\documentclass[11pt]{article}

\usepackage{amsthm, amsmath, amsfonts, amssymb}
\usepackage[numbers]{natbib}
\usepackage[colorlinks, citecolor=blue, urlcolor=blue]{hyperref}
\usepackage{graphicx, xcolor}
\usepackage{dsfont, float, enumitem}
\usepackage[a4paper, margin=1.5 in]{geometry}

\theoremstyle{plain}

\newtheorem{thm}{Theorem}[section]
\newtheorem{lemma}{Lemma}[section]
\newtheorem{cor}{Corollary}[section]
\newtheorem{prop}{Proposition}[section]

{

}
\theoremstyle{plain}
\newtheorem{definition}[thm]{Definition}

\theoremstyle{remark}

\newcommand{\E}{E}
\newcommand{\V}{\text{var}}

\renewcommand{\epsilon}{\varepsilon}

\newcommand{\iid}{\stackrel{\text{i.i.d.}}{\sim}}

\newcommand{\customcitezero}[1]{\href{cite.#1}{\textcolor{blue} {\citeauthor{#1}}} \href{cite.#1}{(\textcolor{blue} {\citeyear{#1}})}}

\newcommand{\customciteone}[1]{(\href{cite.#1}{\textcolor{blue} {\citeauthor{#1},}} \href{cite.#1}{\textcolor{blue} {\citeyear{#1}}})}

\newcommand{\customcitetwo}[2]{(\href{cite.#1}{\textcolor{blue} {\citeauthor{#1},}} \href{cite.#1}{\textcolor{blue} {\citeyear{#1}}}; \href{cite.#2}{\textcolor{blue} {\citeauthor{#2},}} \href{cite.#2}{\textcolor{blue} {\citeyear{#2}}})}

\newcommand{\customcitethree}[3]{(\href{cite.#1}{\textcolor{blue} {\citeauthor{#1},}} \href{cite.#1}{\textcolor{blue} {\citeyear{#1}}}; \href{cite.#2}{\textcolor{blue} {\citeauthor{#2},}} \href{cite.#2}{\textcolor{blue} {\citeyear{#2}}}; \href{cite.#3}{\textcolor{blue} {\citeauthor{#3},}} \href{cite.#3}{\textcolor{blue} {\citeyear{#3}}})}

\newcommand{\customcitefour}[4]{(\href{cite.#1}{\textcolor{blue} {\citeauthor{#1},}} \href{cite.#1}{\textcolor{blue} {\citeyear{#1}}}; \href{cite.#2}{\textcolor{blue} {\citeauthor{#2},}} \href{cite.#2}{\textcolor{blue} {\citeyear{#2}}}; \href{cite.#3}{\textcolor{blue} {\citeauthor{#3},}} \href{cite.#3}{\textcolor{blue} {\citeyear{#3}}}; \href{cite.#4}{\textcolor{blue} {\citeauthor{#4},}} \href{cite.#4}{\textcolor{blue} {\citeyear{#4}}})}

\newcommand{\customcitefive}[5]{(\href{cite.#1}{\textcolor{blue} {\citeauthor{#1},}} \href{cite.#1}{\textcolor{blue} {\citeyear{#1}}}; \href{cite.#2}{\textcolor{blue} {\citeauthor{#2},}} \href{cite.#2}{\textcolor{blue} {\citeyear{#2}}}; \href{cite.#3}{\textcolor{blue} {\citeauthor{#3},}} \href{cite.#3}{\textcolor{blue} {\citeyear{#3}}}; \href{cite.#4}{\textcolor{blue} {\citeauthor{#4},}} \href{cite.#4}{\textcolor{blue} {\citeyear{#4}}}; \href{cite.#5}{\textcolor{blue} {\citeauthor{#5},}} \href{cite.#5}{\textcolor{blue} {\citeyear{#5}}})}

\newcommand{\customref}[1]{\textcolor{pink}{\ref{#1}}}
\newcommand{\customreftwo}[1]{\textcolor{pink}{\ref{#1}}}

\begin{document}
\title{\bf \LARGE
Decomposition-Based Intrinsic Modeling of Shape-Constrained Functional Data}
  \author{\bf \large Poorbita Kundu and Hans-Georg M\"uller \vspace{2mm} \\
  \normalsize \textit{Department of Statistics, University of California, Davis, CA, USA}}
    \date{}
  \maketitle
\bigskip

\begin{abstract}
\noindent Shape-constrained functional data encompass a wide array of application fields, such as activity profiling, growth curves,  healthcare and mortality. Most existing methods for general functional data analysis often ignore that such data are subject to inherent shape constraints, while some specialized techniques rely on strict distributional assumptions. We propose an approach for modeling such data that harnesses the intrinsic geometry of functional trajectories  by decomposing them into size and shape components. We focus on the two most prevalent shape constraints, positivity and monotonicity, and develop individual-level estimators for the size and shape components. Furthermore, we demonstrate the applicability of our approach by conducting subsequent analyses involving Fr\'{e}chet mean and Fr\'{e}chet regression and  establish rates of convergence for the empirical estimators.  Illustrative examples include  simulations and  data applications for activity profiles for Mediterranean fruit flies during their entire lifespan and for data from the Z\"{u}rich longitudinal growth study. 
\end{abstract}


monotonicity, positivity, size-shape decomposition.

\section{Introduction}
\label{sec:intro} 

Functional data containing sample of time-indexed trajectories  are often encountered in real life \customcitethree{ramsay2005principal}{hsing2015theoretical}{wang2016functional}, where such data may be susceptible to shape constraints, with  positivity and monotonicity as the most common constraints.
Examples that we will study in this paper include activity profile data and data on human growth and development. But many other functional data such as  data on  mortality, temperature, biomarkers or longitudinal blood pressure are subject to such constraints. Standard approaches like functional principal component analysis (FPCA) commonly employed for functional data \customcitefive{kleffe1973principal}{dauxois1982asymptotic}{castro1986principal}{hall2006}{chen2015localized} typically do not respect these constraints,  due to the  non-linearity of the space of shape-constrained functional data, as for example linear combinations of positive functions are not necessarily positive.  Consequently, the estimated trajectories obtained through the Karhunen-Loève decomposition \customcitetwo{karhunen1947under}{loeve1978} generally will not  reside in the constrained space. 

For functional data that are constrained to be positive,  a potential remedy is to employ FPCA as a dimension reduction tool on the log-transformed observed data, which are situated in the unconstrained space $L^{2}([0,1])$. Subsequently, we can map the estimates obtained for transformed data  to the original space by exponentiating \customcitetwo{ramsay2005principal}{petersen2016functional}. An analogous transformation approach can be applied for monotone functional data; see Section \ref{section:5}. However, this transformation approach introduces a transformation bias, which can be challenging to address. 

In recent times, the functional data analysis (FDA) literature reflects a growing interest in dealing with shape-constraints in various forms. For example, \customcitezero{ghosal2023shape} imposes shape restriction on regression coefficient functions; on the other hand, shape-constrained generalized additive models \customcitetwo{pya2015shape}{chen2016generalized} impose shape restrictions on components of additive prediction function. However, a notable gap persists in addressing shape constraints that are imposed directly on the underlying stochastic process --- a consideration that becomes particularly relevant when dealing with shape-constrained functional data. 

To address these limitations, we propose a decomposition-based approach for shape-constrained functional data residing in non-linear space. In particular, the decomposition approach is used to intrinsically model shape-constrained functional data, mitigating the need for transformation, while ensuring compliance to the inherent shape restrictions. The specific definition and interpretation of the decomposition components depend on the nature of the constraint. In general, the size component represents the overall amplitude of the trajectory, while the shape component captures temporal (or age-related) variations and preserves the shape constraint by construction. This individual-level size-shape decomposition not only offers flexibility but also explores the intrinsic geometry of functional data, enhancing interpretability. \customcitezero{gajardo2021cox} uses a related decomposition for the intensity function in Cox point process regression.  

Our proposed approach is novel and does not rely on narrow distributional or structural assumptions, making it applicable to all positive or monotone constrained functional data. It is nonparametric, does not involve the use of functional principal components and respects the shape-constraints in the data, without introducing data distortion or loss of interpretability. We establish uniform rates of convergence for the proposed estimates, uniform over subjects as well as the time domain. In the presence of covariates, the proposed approach utilizes Fr\'{e}chet regression \customciteone{petersen2019frechet} for the decomposition components and it will be  illustrated  with data on medfly lifetime activity profiles and data from the Z\"urich longitudinal growth study.

The rest of the paper is organized as follows. Section \customreftwo{section:2} introduces the decomposition approach for shape-constrained functional data, which is tied  to local and global Fr\'{e}chet regression for the size-shape decomposition components in Section   \customreftwo{section:3}.  In Section \customreftwo{section:4}, we introduce the proposed estimates and establish their rates of convergence. Simulation studies and data applications are discussed in Sections \customreftwo{section:5} and \customreftwo{section:6} respectively, followed by a brief discussion in Section \customreftwo{section:7}. Proofs, auxiliary results and additional data illustrations are provided in the Appendix. 

\section{Modeling Shape-Constrained Functional Data} \label{section:2} 

\subsection{Positive functional data}
\label{subsection:2.1}
Let $Y: \mathcal{T} \rightarrow \mathcal{S} \subset (0, \infty)$ be a strictly positive-valued stochastic process defined on a compact time domain $\mathcal{T}$. Without loss of generality, we assume $\mathcal{T} = [0,1]$. Denoting the space of strictly positive functions by $$\Omega=\{ Y\in  L^{2}([0,1]):\, Y(t)>0, \: t \in [0,1]\},$$ for each $Y \in \Omega$, we consider the decomposition into two components, namely,
\begin{definition}
   \textit{Size component:} A scalar $\tau$ denoting the overall magnitude of the positive function, defined as $\tau = \int_{0}^{1} Y(t) d t .$
\end{definition}

\begin{definition}
     \textit{Shape component:} A density function $ f:[0, 1] \rightarrow (0, \infty)$ preserving the positivity constraint, defined as $f(t):=\frac{Y(t)}{\tau}$.
\end{definition}

We use the symbols $\Omega_{\mathcal{T}}$ and $\Omega_{\mathcal{F}}$ to denote the spaces for size and shape components, respectively. In particular, $\Omega_{\mathcal{T}}=\mathbb{R}^{+}$, while $\Omega_{\mathcal{F}}$ represents the space of density functions associated with probability measures on $[0,1]$, which can also be represented in terms of the corresponding quantile functions. Specifically, we use $\mathcal{Q}\bigl(\Omega_\mathcal{F}\bigr)$ to denote the space of quantile functions for $\Omega_\mathcal{F}$, where each shape function $f \in \Omega_\mathcal{F}$ has a unique associated probability measure $\nu(f)$ that can be quantified in terms of its cumulative distribution function $\nu(F)$ or its quantile function $\nu(Q)$. The decomposition $Y = \tau f$ naturally defines a one-to-one correspondence between $Y$ and $(\tau, f)$ through a $1:1$ map $\Psi: \Omega \rightarrow \Omega_{\mathcal{T}} \times \Omega_{\mathcal{F}}$. Endowing $\Omega_{\mathcal{T}}$ and $\Omega_{\mathcal{F}}$ with metrics $d_{\mathcal{T}}$ and $d_{\mathcal{F}}$, respectively, $(\Omega, d)$ can be regarded as a product metric space $\bigl(\Omega_{\mathcal{T}}, d_{\mathcal{T}}\bigr) \times\bigl(\Omega_{\mathcal{F}}, d_{\mathcal{F}}\bigr)$, with metric $d$ defined as
\begin{align}
\label{metric:d:on:Omega}
  d\bigl(Y_1,Y_2\bigr) = d\bigl(\Psi^{-1}\bigl(\tau_{1}, f_{1}\bigr),\Psi^{-1}\bigl(\tau_{2}, f_{2}\bigr)\bigr):=\sqrt{d_{\mathcal{T}}^{2}\bigl(\tau_{1}, \tau_{2}\bigr)+d_{\mathcal{F}}^{2}\bigl(f_{1}, f_{2}\bigr)} \ ,  
\end{align}
\noindent where $Y_1, Y_2 \in \Omega$. In some cases one can also consider more general metrics  given by $d(Y_1,Y_2)= \sqrt{w_1 \ d_{\mathcal{T}}^{2}\bigl(\tau_{1}, \tau_{2}\bigr)+ w_2 \ d_{\mathcal{F}}^{2}\bigl(f_{1}, f_{2}\bigr)}$ with weights $w_1, w_2 \geq 0$. The theoretical results in Section \customref{section:4} can be easily extended to this general case.  

We select the Euclidean metric $d_{E}$ as a natural choice for $d_{\mathcal{T}}$ and the 2-Wasserstein metric for $d_{\mathcal{F}}$. While  other choices might also be of interest,  the Wasserstein metric $d_{W}$  has proved to be useful in practical applications involving samples of distributions \customcitetwo{bolstad2003comparison}{zhang2011functional}. Let $\nu(f)$, $\nu(Q)$, $\nu(F)$ denote the probability measure  associated with a given probability density function $f$, quantile function $Q$, and cumulative distribution function $F$. For two probability measures $\nu_1$ and $\nu_2$ with quantile functions $Q_{\nu_1}$ and $Q_{\nu_2}$, the 2-Wasserstein distance for univariate distributions is $$d_{W}^{2}\bigl(\nu_1,\nu_2\bigr) = \int_{0}^{1}\bigl(Q_{\nu_1}(t)-Q_{\nu_2}(t)\bigr)^{2} d t.$$ 

For any two densities $f_1, f_2 \in \Omega_{\mathcal{F}}$, we write $d_{W}^{2}\bigl(f_{1}, f_{2}\bigr) := d_{W}^{2}\bigl(\nu(f_1),\nu(f_2)\bigr)$ and analogously $d_{W}^{2}\bigl(F_{1}, F_{2}\bigr) := d_{W}^{2}\bigl(\nu(F_1),\nu(F_2)\bigr)$ for any two cumulative distribution functions $F_1$ and $F_2$.  
The Fr\'echet mean of $Y$ is defined as $Y_{\oplus}=\underset{\omega \in \Omega}{\operatorname{argmin}} \ V(\omega)$, which denotes the  potentially multiple minimizers of the Fr\'{e}chet function $$V(\omega):= \mathbb{E}\bigl(d^{2}\bigl(Y, \omega\bigr)\bigr) = \mathbb{E}\bigl(d_{E}^{2}\bigl(\tau, \tau(\omega) \bigr)\bigr)+\mathbb{E}\bigl(d_{W}^{2}\bigl(f, f(\omega)\bigr)\bigr),$$ where $\bigl(\tau(\omega), f(\omega)\bigr) = \Psi\bigl(\omega\bigr)$ and $(\tau, f) = \Psi\bigl(Y\bigr)$. Thus, $$\Psi\bigl(Y_{\oplus}\bigr) = \underset{\tau(\omega) \in \Omega_{\mathcal{T}}, f(\omega) \in \Omega_{\mathcal{F}}
}{\operatorname{argmin}}\bigl( \mathbb{E}\bigl[d_{E}^{2}\bigl(\tau, \tau(\omega) \bigr)\bigr]+\mathbb{E}\bigl[d_{W}^{2}\bigl(f, f(\omega)\bigr)\bigr]\bigr)
= \bigl(\tau_{\oplus} , f_{\oplus} \bigr),$$ where $\tau_{\oplus}=\underset{\tau(\omega) \in \Omega_{\mathcal{T}}}{\operatorname{argmin} } \ \mathbb{E}\bigl(d_{E}^{2}\bigl(\tau, \tau(\omega)\bigr) \bigr)$ and $
f_{\oplus}=\underset{f(\omega) \in \Omega_{\mathcal{F}}}{\operatorname{argmin} } \ \mathbb{E}\bigl(d_{W}^{2}\bigl(f, f(\omega)\bigr)\bigr)$ is the density corresponding to the quantile function $Q_{\oplus} =\underset{Q(\omega) \in {\mathcal{Q}}(\Omega_{\mathcal{F}})}{\operatorname{argmin} } \mathbb{E}\bigl(d_{L^2}^{2}\bigl(Q, Q(\omega)\bigr)\bigr)$. 
The optimization problem is thus separable and has a unique  solution $Y_{\oplus}=\Psi^{-1}\bigl(\bigl(\tau_{\oplus}, f_{\oplus}\bigr)\bigr) = \tau_{\oplus} \cdot f_{\oplus}.$

\subsection{Monotone functional data}
\label{mono:decomposition}
Let $Y:[0,1] \rightarrow \mathcal{R}$ be a stochastic process with smooth and srictly monotone increasing (non-constant) trajectories. We write  $\Omega \subset L^2([0,1])$ to denote the corresponding function space. We focus here on monotone increasing functional data 
and our considerations apply analogously to processes with monotone decreasing trajectories. We decompose each monotone trajectory $Y \in \Omega$ into
\begin{definition}
    \textit{Size component:} A two-dimensional vector $\xi=\bigl(\rho,\lambda\bigr)$ denoting the range and the minimum value of the monotone function, defined as $\rho = Y(1) - Y(0) > 0 \ \text{and }  \lambda = Y(0).$ 
\end{definition}

\begin{definition}
    \textit{Shape component:} A cumulative distribution function $ F:[0, 1] \rightarrow [0,1]$ preserving the monotonicity constraint, 
    defined as $F(t):= \frac{Y(t) - \lambda}{\rho}.$
\end{definition}
\noindent With similar notations as in the case of positive functions, $Y=\lambda + \rho F$ defines a correspondence between $Y$ and $(\xi, F)=(\rho,\lambda,F)$ through the $1:1$ map $\Psi: \Omega \rightarrow \Omega_{\mathcal{T}} \times \Omega_{\mathcal{F}}$. Again, if we endow spaces $\Omega_{\mathcal{T}}$ and $\Omega_{\mathcal{F}}$ with metrics $d_{\mathcal{T}}$ and $d_{\mathcal{F}}$ respectively, $(\Omega, d)$ can be regarded as a product metric space $\bigl(\Omega_{\mathcal{T}}, d_{\mathcal{T}}\bigr) \times\bigl(\Omega_{\mathcal{F}}, d_{\mathcal{F}}\bigr)$. With metric choices $d_{\mathcal{T}}\bigl(\xi_{1}, \xi_{2}\bigr):=d_{E}\bigl(\xi_{1}, \xi_{2}\bigr) = \sqrt{(\rho_1-\rho_2)^2 +(\lambda_1-\lambda_2)^2}$ for $\xi_1, \xi_2 \in \Omega_{\mathcal{T}}$ and $d_{\mathcal{F}}\bigl(F_{1}, F_{2}\bigr):=d_{\mathbb{L}^2([0,1])}\bigl(Q_{1}, Q_{2}\bigr) = d_W\bigl(\nu(F_1),\nu(F_2)\bigr)$, the metric $d$ in $\Omega$ is defined as
\begin{align}
\label{metric:d:on:Omega:mono}
  d\bigl(Y_1,Y_2\bigr) & = \sqrt{d_{E}^{2}\bigl(\xi_{1}, \xi_{2}\bigr)+d_{W}^{2}\bigl(F_{1}, F_{2}\bigr)} \nonumber\\
  & = \sqrt{ (\rho_1-\rho_2)^2 +(\lambda_1-\lambda_2)^2 + d^2_{\mathbb{L}^2([0,1])}\bigl(Q_{1}, Q_{2}\bigr) }.  
\end{align}

\noindent Fr\'{e}chet means are  $\xi_{\oplus}=\underset{\xi(\omega) \in \Omega_{\mathcal{T}}}{\operatorname{argmin} } \ \mathbb{E}\bigl(d_E^{2}\bigl(\xi, \xi(\omega)\bigr) \bigr) := \bigl(\rho_{\oplus}, \lambda_{\oplus}\bigr)$ and\\  $F_{\oplus}=\underset{F(\omega) \in \Omega_{\mathcal{F}}}{\operatorname{argmin} } \ \mathbb{E}\bigl(d_W^{2} \bigr. $
$\bigl. \bigl(F, F(\omega)\bigr)\bigr)$,  with $Q_{\oplus}= F^{-1}_{\oplus} = \underset{Q(\omega) \in {\mathcal{Q}}(\Omega_{\mathcal{F}})}{\operatorname{argmin} } \mathbb{E} \bigl(d_{L^2}^{2}\bigl(Q, Q(\omega)\bigr)\bigr)$.\\ Combining these minimizers, we obtain $Y_{\oplus} = \lambda_{\oplus} + \rho_{\oplus} \cdot F_{\oplus}$.

\section{Functional regression under shape-constraints}
\label{section:3} 
We now extend our discussion beyond Fr\'{e}chet means and regress the decomposition components  on Euclidean predictors. Considering positive functional data, the local Fr\'{e}chet regression function of $Y$ on $X \in \mathcal{X} \subset \mathcal{R}^p$ is 
\begin{align}
\label{Y:plus:x}
    Y_\oplus(x) & = \underset{Y_0 \in \Omega}{\operatorname{argmin}} \ \mathbb{E}\bigl(d^2\bigl(Y, Y_0\bigr) \big\vert X=x\bigr) \nonumber \\
    & = \underset{(\tau_0,f_0) \in \Omega_{\mathcal{T}} \times \Omega_{\mathcal{F}}}{\operatorname{argmin}} \mathbb{E}\bigl(d_{E}^{2}\bigl(\tau, \tau_{0}\bigr) + d_{W}^{2}\bigl(f, f_0\bigr) \big\vert X=x\bigr) \nonumber \\
    & = \Psi^{-1}(\tau_{\oplus}(x), f_{\oplus}(x)) = \tau_{\oplus}(x) \cdot f_{\oplus}(x), 
\end{align}
where $\tau_{\oplus}(x)$ and $f_{\oplus}(x)$ are defined as
\begin{align}
    \label{tau:f:Y:plus:x}
     \tau_{\oplus}(x) & = \underset{\tau_0 \in \Omega_\mathcal{T}}{\operatorname{argmin}} \ \mathbb{E}\bigl[d_E^2\bigl(\tau, \tau_0 \bigr) \big\vert X=x\bigr] =  \mathbb{E}\bigl[\tau \big\vert X=x\bigr] \text{ and } \nonumber \\
    f_{\oplus}(x) & = \underset{f_0 \in \Omega_\mathcal{F}}{\operatorname{argmin} } \ \mathbb{E}\bigl[d_W^2\bigl(f, f_0\bigr) \big\vert X=x\bigr].
\end{align}

\noindent Extending this approach  to monotone functional data, analogous to $\tau_{\oplus}(x)$ in (\customref{tau:f:Y:plus:x}), we define  $\xi_{\oplus}(x) = \bigl(\rho_{\oplus}(x), \lambda_{\oplus}(x)\bigr)$, where $\rho_{\oplus}(x) = \mathbb{E}\bigl[\rho \big\vert X=x\bigr]$ and $\lambda_{\oplus}(x) = \mathbb{E}\bigl[\lambda \big\vert X=x\bigr] $. Analogous to $f_{\oplus}(x)$ in (\customref{tau:f:Y:plus:x}), the localized Fr\'{e}chet mean for shape is $$F_{\oplus}(x)=\underset{F(\omega) \in \Omega_\mathcal{F}}{\operatorname{argmin}} \ \mathbb{E}\bigl(d_{W}^2\bigl(F, F(\omega)\bigr) \big\vert X=x\bigr)$$ and can alternatively be obtained through quantile function representations  $$Q_{\oplus}(x) = F^{-1}_{\oplus}(x)= \underset{Q(\omega) \in \mathcal{Q}\bigl(\Omega_\mathcal{F}\bigr)}{\operatorname{argmin}} \mathbb{E}\bigl(d_{L^2([0,1])}^2\bigl(Q, Q(\omega)\bigr) \big\vert X=x\bigr).$$ Using $Y = \lambda + \rho F$ in (\customref{metric:d:on:Omega:mono}), 
\begin{align}
\label{Y:f:tau:plus:mono}
Y_{\oplus}(x)= \lambda_{\oplus}(x) + \rho_{\oplus}(x) F_{\oplus}(x).
\end{align}

\noindent To facilitate  inclusion of  categorical predictors into our model, we use a global Fr\'{e}chet regression function $Y_{\mathrm{G_\oplus} }(x) := \underset{w \in \Omega}{\operatorname{argmin} } \ \mathbb{E}\bigl(s(X, x)  d^2(Y, w)\bigr)$, with $s(X, x) \\ :=1+(X-\mu)^T \Sigma^{-1}(x-\mu)$. We assume $\mu:=\mathbb{E}(X)$ and $\Sigma:=\operatorname{Var}(X)$ exist, with positive definite $\Sigma$.
The key idea involves characterization of multiple linear regression as a weighted sum of the responses, which can then be generalized to the case of weighted  Fr\'{e}chet means \customciteone{petersen2019frechet}, wherein the global weights $s(X, x)$ do not depend on tuning parameters, unlike local methods. Using metric decomposition for positive functional $Y$, 
\begin{align}
\label{tau:f:Gplus:defintion}
& Y_{\mathrm{G_\oplus}}(x)=\tau_{\mathrm{G_\oplus}}(x) \cdot f_{\mathrm{G_\oplus}}(x), \text{ where } \tau_{\mathrm{G_\oplus}}(x)= \underset{\tau_0 \in \Omega_{\mathcal{T}}}{\operatorname{argmin} } \ \mathbb{E}\bigl(s(X, x) \ d_{E}^2\bigl(\tau, \tau_0\bigr)\bigr) \text{ and } \nonumber \\
& f_{\mathrm{G_\oplus}}(x)=\underset{f_0 \in \Omega_{\mathcal{F}}}{\operatorname{argmin} } \ \mathbb{E}\bigl(s(X, x) \ d_{W}^2\bigl(f, f_0\bigr)\bigr).
\end{align}
\noindent Analogous to $\tau_{\mathrm{G_\oplus}}(x)$ in (\customref{tau:f:Gplus:defintion}) for monotone functional data, $$\xi_{G_\oplus}(x) = \underset{\xi_0 \in \Omega_{\mathcal{T}} \subset \mathcal{R}^2}{\operatorname{argmin} } \ \mathbb{E}\bigl(s(X, x) \ d_{E}^2\bigl(\xi, \xi_0\bigr)\bigr) \newline = \bigl(\rho_{G_\oplus}(x), \lambda_{G_\oplus}(x)\bigr).$$  For shape, ${F}_{G_\oplus}(x)=\underset{F(\omega) \in \Omega_{\mathcal{F}}}{\operatorname{argmin}} \ \mathbb{E}(s(X, x) \ d_{W}^2(F, F(\omega))$ with $${Q}_{G_\oplus}(\cdot, x)= {F}^{-1}_{G_\oplus}(x) = \underset{Q(\omega) \in \mathcal{Q}\bigl(\Omega_\mathcal{F}\bigr)}{\operatorname{argmin}} \mathbb{E}\bigl(s(X, x) \ d_{L^2([0,1])}^2\bigl(Q, Q(\omega)\bigr)\bigr).$$ Using $Y= \lambda + \rho F$ in (\customref{metric:d:on:Omega:mono}), the final representation of the fitted monotone function, regressed on the predictor level $X$, is as follows, 
\begin{align}
\label{Y:f:tau:G:plus:mono}
Y_{G_\oplus}(x)= \lambda_{G_\oplus}(x) + \rho_{G_\oplus}(x) F_{G_\oplus}(x).
\end{align}

\section{Estimation and Convergence}
\label{section:4}
\subsection{Recovery of individual trajectories}
\label{section:4.1}
Recall that $Y: \mathcal{T} \rightarrow \mathcal{S}$ are the underlying smooth random trajectories. We observe $n$ data trajectories $\bigl\{Z_i \bigr\}_{i=1}^{n}$ at $N_i$ equidistant time points $\{t_{ij}\}_{j=1}^{N_i}$ and assume that the available measurements are noisy. The data model is then $Z_{i j}= Z_i\bigl(t_{ij}\bigr)=Y_i\bigl(t_{ij}\bigr)+\varepsilon_{i j}$, $j = 1, \dots, N_i$.  
A first step is to recover the individual-level true (unobserved) trajectories $Y_i$ nonparametrically  as follows.  We divide $\mathcal{T} = [0,1]$ into $L$ bins of equal width. To establish a uniform rate of convergence over all the subjects, we use a common bin size $\gamma$, resulting in $L = \gamma^{-1}$ bins $\bigl\{B_\ell\bigr\}_{\ell=1}^L$, where $B_\ell = [(\ell-1)\gamma, \ell\gamma]$. Then the underlying trajectory $Y_i$ is estimated by
\begin{align}
\label{Est:Yi:Step}
   \widehat{Y}_i(t) =  \frac{\sum_{j=1}^{N_i} \ Z_{ij} \mathds{1}_{\{t_{ij} \in B_\ell\}}}{\sum_{j=1}^{N_i} \ \mathds{1}_{\{t_{ij} \in B_\ell\}}}, \ \text{for} \ t \in B_\ell.
\end{align}

To establish  rates of convergence, we require a few additional assumptions. 
\begin{enumerate}[label = (A\arabic*), series = fregStrg, start = 1]
    \item \label{ass:A1} The measurement noise $\epsilon$ is independent of the underlying stochastic process $Y$. We assume $\varepsilon_{i j} \iid \varepsilon$ with $\E(\varepsilon) = 0$ and $\V(\varepsilon) = \sigma^2<\infty$. 
    \item \label{ass:A2} There exists a sequence of positive integers $N:=N(n)$ such that $N < \underset{1 \leq i \leq n}{\inf} N_i$ and $\log n \big/ N \rightarrow 0$ as $n \rightarrow \infty$. 
    \item \label{ass:A3} $Y(t)$ is twice continuously differentiable and there exist real constants $c_1, c_2 > 0$ such that  $|Y^{'}(0)|\leq c_1$ and $\|Y^{''}\|_{\infty} \leq c_2$ with probability 1.
\end{enumerate} 

\noindent Assumption \customreftwo{ass:A1} is a standard  assumption when functional data are measured with noise.  Assumption \customreftwo{ass:A2} provides a uniform lower bound on the number of observations for each subject, used to obtain rates uniform over subjects. 
Assumption \customreftwo{ass:A3} imposes a smoothness constraint on the trajectories. With these assumptions in place, we  obtain a rate of convergence $O_{p}\bigl((\log n\big/N)^{1/3}\bigr)$, where this rate is uniform over both the time domain and also all the subjects (see Appendix \ref{appA}, Theorem \ref{Theorem:Step:Est:Yi:hat}).

\subsection{Estimation and theory}
\label{theory:posit:mono}
We use the individual level estimated trajectories defined in (\customref{Est:Yi:Step}) to obtain the following estimates of the latent shape and size components for positive functional data,
\begin{align}
\label{Pos:decomp:est}
    \widehat{\tau}_i := \frac{1}{N} \sum_{j=1}^{N} Z_{ij} \quad \text {and} \quad  \widehat{f}_i(t):=\frac{\widehat{Y}_i(t)}{\int_{0}^{1} \widehat{Y}_i(t) dt} , \quad \text {for all } t \in [0,1]  \quad \hspace{-2mm} \text{and each } i.
\end{align}

\noindent We additionally require  assumption \customreftwo{ass:S0} below to ensure that the trajectories are  bounded away from zero, to guarantee that the shape estimates 
 are well-defined. 
\begin{enumerate}[label = (S\arabic*), series = fregStrg, start = 0]
    \item \label{ass:S0} There exist constants $0 < \kappa_0, \kappa_1 < \infty$ such that realizations of the underlying stochastic process $Y$ satisfy $\kappa_0 \leq \inf_{t \in [0, 1]} {Y(t)}$ and $ \|Y\|_{\infty} \leq \kappa_1$.
\end{enumerate}
\vspace{2mm}

\begin{thm}
\label{Theorem 2:uniform:tau:shape}
Under  \customreftwo{ass:A1}$-$\customreftwo{ass:A3} the  size estimates  $\widehat{\tau}_i$ defined in (\customref{Pos:decomp:est}) satisfy
\begin{align*}     
        n^{-1} \sum_{i=1}^{n}  \bigl|\widehat{\tau}_i-\tau_i\bigr|=O_p\bigl(N^{-1/2}\bigr) \text{ and } \sup_{1 \leq i \leq n}  \bigl|\widehat{\tau}_i-\tau_i\bigr|=O_p\bigl((\log n\big/N)^{1/2}\bigr).
    \end{align*}
 Under \customreftwo{ass:S0}, with  bin sizes $\gamma  \sim \bigl(\log n\big/N\bigr)^{1/3}$, the shape estimates $\widehat{f}_i(t)$ in (\customref{Pos:decomp:est}) satisfy 
   \begin{align*}
    \sup_{1 \leq i \leq n}  \sup_{t \in [0, 1]} \bigl|\widehat{f}_i(t)-f_i(t)\bigr| &=  O_p\bigl((\log n\big/N)^{1/3}\bigr) \ \text{and} 
    \\ 
    \sup_{1 \leq i \leq n} d_{W} \bigl(\widehat{f}_i,f_i\bigr)&=O_p\bigl((\log n\big/N)^{1/6}\bigr).
    \end{align*}
\end{thm}

\noindent The first part of Theorem \customreftwo{Theorem 2:uniform:tau:shape} provides  convergence rates for the size components, both on  average and uniformly over subjects. The second part illustrates individual-level (uniform) convergence for the shape components, again uniformly over subjects. 

Concerning estimation of the corresponding Fr\'{e}chet means defined in Section \customref{subsection:2.1}, $\tau_{\oplus}$ is unique by definition, and Lemma \ref{Lemma 1} (in Appendix \ref{appA}) guarantees the unique existence of $f_{\oplus}$. Therefore, the minimizer $Y_{\oplus}$ exists uniquely. The estimate for $\tau_{\oplus}$ is given by $\widehat{\tau}_{\oplus} = n^{-1} \sum_{i=1}^{n} \widehat{\tau}_i$ and $f_{\oplus}$ is estimated by the density $\widehat{f}_{\oplus}$, corresponding to the quantile function
 $\widehat{Q}_{\oplus} = \underset{q \in {\mathcal{Q}}(\Omega_{\mathcal{F}})}{\operatorname{argmin} } \bigl\| q - n^{-1} \sum_{i=1}^{n} \widehat{Q}_i \bigr\|_{L^2([0,1])}^2$.

 \begin{cor}
    \label{Corollary:Tau:f:Y:plus:EQUI}
     Under assumptions \customreftwo{ass:A1}$-$\customreftwo{ass:A3}, 
     $$\bigl|\widehat{\tau}_{\oplus} - \tau_{\oplus}\bigr|= O_p\bigl({N}^{-1/2} + n^{-1/2} \bigr).$$
     Under  \customreftwo{ass:S0}, with $\gamma \sim (\log n\big/N)^{1/3}$, furthermore  
     \begin{align*}
     d_W\bigl(f_{\oplus}, \widehat{f}_{\oplus}\bigr)  & =  O_p\bigl((\log n\big/N)^{1/6} + n^{-1/2}\bigr), 
    \text{ so that }  \\
     d\bigl(Y_{\oplus}, \widehat{Y}_{\oplus}\bigr) & = O_p\bigl((\log n\big/N)^{1/6} + n^{-1/2}\bigr).
    \end{align*}
\end{cor}

To facilitate the inclusion of  categorical predictors in the regression model, we primarily focus on estimating the global Fr\'{e}chet regression functions $\tau_{G_\oplus}(x)$ and $f_{G_\oplus}(x)$. We obtain regression estimates adopting the empirical global weights in \customcitezero{petersen2019frechet}, defined as $s_{i n}(x):=1+ (X_i-\bar{X})^T \hat{\Sigma}^{-1}(x-\bar{X})$, where $\bar{X}:=n^{-1} \sum_{i=1}^n X_i$ and $\hat{\Sigma}:=n^{-1} \sum_{i=1}^n (X_i-\bar{X})(X_i-\bar{X})^T$. The resulting estimate for the scalar size factor $\tau_{G_\oplus}(x)$ 
is the usual least squares estimator $\widehat{\tau}_{G_\oplus}(x) = \operatorname{max} \{\kappa_0, n^{-1} \sum_{i=1}^n s_{i n}(x) \widehat{\tau}_i \}$, constrained to be positive. To obtain the resulting regression for the  shape function on Euclidean predictors, we use quantile functions $Q_i$ and $\widehat{Q}_i$ of  $f_i$ and $\widehat{f}_i$. Deploying empirical global weights $s_{i n}(x)$ then leads to the global Fr\'echet regression estimator 
\begin{equation}
\label{Q:hat:Gplus:x:star}
    \widehat{Q}_{G_\oplus}(\cdot, x) = \underset{Q_0 \in \mathcal{Q}(\Omega_\mathcal{F})}{\operatorname{argmin}} \bigl\|Q_0 - n^{-1} \sum_{i=1}^n s_{i n}(x) \widehat{Q}_i\bigr\|^2_{L^2([0,1])}, 
\end{equation}
with corresponding  density function $\widehat{f}_{G_\oplus}(x)$.
Then (\customref{tau:f:Gplus:defintion}) leads to the final estimate $\widehat{Y}_{G_\oplus}(x) = \widehat{f}_{G_\oplus}(x) \cdot \widehat{\tau}_{G_\oplus}(x)$ of $Y_{G_\oplus}(x)$. 
\vspace{2mm}

\begin{thm}
\label{Theorem 8B:Global:regg:tau:f:Gplus:x}
Under assumptions \customreftwo{ass:A1}$-$\customreftwo{ass:A3}, it holds that
\begin{align*}
\bigl|\tau_{G_{\oplus}}(x) - \widehat{\tau}_{G_{\oplus}}(x)\bigr| = O_p\bigl(N^{-1/2}+n^{-1/2}\bigr).
\end{align*}
\text{Additionally under assumption \customreftwo{ass:S0} and with bin size $\gamma  \sim (\log n\big/N)^{1/3}$,}
\begin{align*}
& d_W\bigl(f_{G_{\oplus}}(x), \widehat{f}_{G_{\oplus}}(x)\bigr)=O_p\bigl((\log n\big/N)^{1/6} + n^{-1/2}\bigr).
\end{align*}
\end{thm}

\noindent Using Theorem \customref{Theorem 8B:Global:regg:tau:f:Gplus:x}, it is easy to see that $\widehat{Y}_{G_{\oplus}}$ enjoys the same convergence rate as $\widehat{f}_{G_{\oplus}}$. 
\vspace{4mm}

The remainder of this section pertains to monotone functional data. Based on $\widehat{Y}_i$, the individual level estimates for size and shape components are   
\begin{align}
\label{Mono:decomp:est}
    & \widehat{\xi}_i := \bigl(\widehat{\rho}_i, \widehat{\lambda}_i\bigr), \text{ with }  \widehat{\rho}_i = \widehat{Y}_i(1) -  \widehat{Y}_i(0) , \ \widehat{\lambda}_i = \widehat{Y}_i(0)  \ \ \text{and} \nonumber\\
    &  \widehat{F}_i(t):= \frac{\widehat{Y}_i(t) - \widehat{\lambda}_i}{\widehat{\rho}_i} 
    , \ \ \text{for all } t \in [0,1].
\end{align}

\noindent Under suitable assumptions, one can establish uniform rates of convergence for the decomposition components associated with monotone trajectories, similar to the case of positive trajectories.  For the size component $\xi$, one can obtain a  $O_p\bigl((\log n\big/N)^{1/3}\bigr)$ (uniform) convergence rate and for  the shape component $F$ the same  convergence rate as for  the case of positive trajectories. Following arguments similar to positive functional data,  we then obtain  convergence rates for Fr\'{e}chet means and Fr\'{e}chet regression functions. The results are provided in Appendix \ref{supp:theory:mono}.

\section{Simulation Study} 
\label{section:5}

We demonstrate the applicability of our proposed method in recovering the latent decomposition components through simulation studies conducted in a number of setups. To generate positive functional trajectories, we adopt a model that adheres to the positivity constraint. We simulate the sample path $Y_i$ on a regular time grid $\{t_{ij}\}_{j=1}^{N} \subset \mathcal{T}$, for $i=1, \dots, n$. The underlying trajectory is modeled as $Y_{ij} := Y_i(t_{ij}) = \tau_i \cdot f_i(t_{ij})$, where the shape component $f_i$ denotes the density of $\mathcal{N}(\mu_i,\sigma^2)$ truncated on time interval $\mathcal{T}$, the size component $\tau_i \iid \mathcal{U}(a_\tau, b_\tau)$, and $\mu_i \iid \mathcal{U}(a_\mu, b_\mu)$. We use $a_\tau > 0$ to ensure the positivity constraint of the underlying trajectories. A similar approach is used for simulating monotone functional data. Instead of $\tau_i$, we now generate the size component $\xi_i = (\rho_i, \lambda_i)$ with $\rho_i \iid \mathcal{U}(a_\rho, b_\rho)$ and $\lambda_i \iid \mathcal{U}(a_\lambda, b_\lambda)$. We then simulate the sample path $Y_i$ at $\{t_{ij}\}_{j=1}^{N}$ as $Y_{ij} := Y_i(t_{ij}) = \rho_i \cdot F_i(t_{ij}) + \lambda_i$, where the shape component $F_i$ is the cumulative distribution function of $\mathcal{N}(\mu_i,\sigma^2)$ truncated on $\mathcal{T}$, which guarantees the monotonicity of the sample path $Y_i$. To mimic real-life data and illustrate the effect of noise, we contaminate the true trajectories with white noise $\epsilon_{ij} \iid \mathcal{N}(0,\nu_0^2)$, resulting in observed values $z_{ij} = Y_{ij} + \epsilon_{ij}$. Consequently, our simulated dataset on the time domain $\mathcal{T}$ comprises the pairs $(t_{ij}, z_{ij})$, where $j = 1, \dots, N$ and $i = 1, \dots, n$.  
\vspace{1mm}

We studied the performance of the proposed approach for  $N = 100, 200, 500,$ $ 1000$ observations per subject and noise levels $\nu_0 = 0.05, 0.1, 0.15$, setting $a_\tau = 0,$ $ b_\tau = 2, a_\mu = 0, b_\mu = 1, \sigma = 1, a_\rho = 0, b_\rho = 4, a_\lambda = -2, b_\lambda = 2, \text{ and time interval}$ $\mathcal{T} = [0, 1]$. For each $(N, \nu_0)$, the simulation study was performed for $n=500$ independent units and was repeated $B=1000$ times. For the $b$th replicate, the error in estimating the quantity of interest was measured by root-mean-square error (RMSE) based on the corresponding metric defined in Sections \ref{subsection:2.1} and \ref{mono:decomposition}. Note that we are interested in estimating the underlying trajectory, size and shape components. For positive functional data, for the $b$th replicate,  $\quad {\text{RMSE}}_b^Y := {\bigl(n^{-1} \sum_{i=1}^{n} d^2(Y_i,\widehat{Y_i})\bigr)}^{1/2}, \qquad  {\text{RMSE}}_b^\tau := {\bigl(n^{-1} \sum_{i=1}^{n} d_{E}^2(\tau_i,\widehat{\tau_i})\bigr)}^{1/2},$ \\ $\text{and} \qquad {\text{RMSE}}_b^f $ $:= {\bigl(n^{-1} \sum_{i=1}^{n} d_{W}^2(f_i,\widehat{f_i})\bigr)}^{1/2}$. Analogously, for monotone 
data,  $\qquad {\text{RMSE}}_b^Y $ $:= {\bigl(n^{-1} \sum_{i=1}^{n} d^2(Y_i,\widehat{Y_i})\bigr)}^{1/2},$ $\qquad {\text{RMSE}}_b^\xi := $  ${\bigl(n^{-1} \sum_{i=1}^{n} d_{E}^2(\xi_i,\widehat{\xi_i})\bigr)}^{1/2},$ $\text{and} \qquad {\text{RMSE}}_b^F := {\bigl(n^{-1} \sum_{i=1}^{n} d_{W}^2(F_i,\widehat{F_i})\bigr)}^{1/2}$. 

\begin{figure}[ht]
    \begin{center}
     \includegraphics[width=0.97\linewidth, height = 12cm]{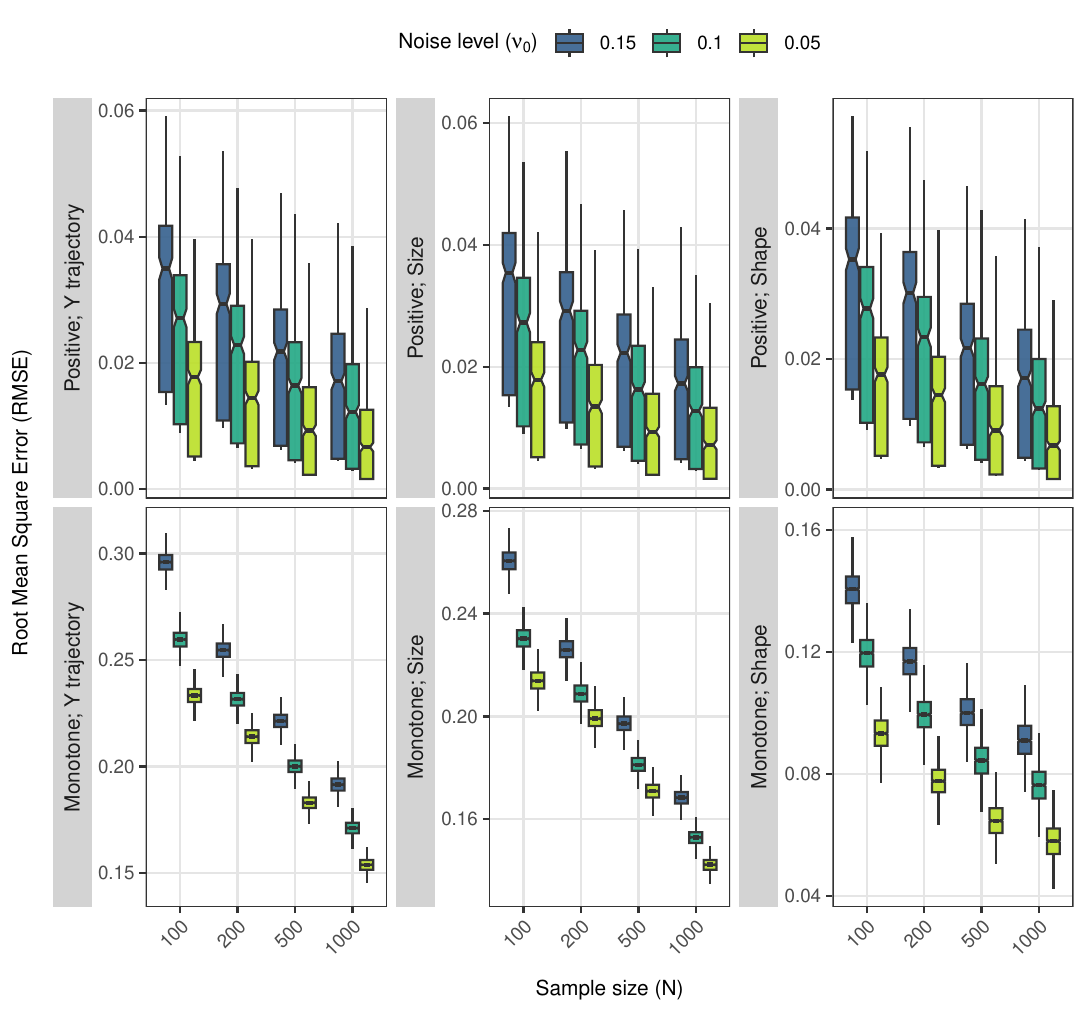}
    \caption{RMSE for the estimation of underlying trajectories $Y_i$, with the proposed size-shape decomposition method on the number of observations $N$ available per subject and the noise variance $\nu^2_0$, both for positive and monotone functional data.}
    \label{fig:sim-posit-mono}
    \end{center}
\end{figure}

As illustrated in Figure \customref{fig:sim-posit-mono},  estimation errors are overall quite small and reflect the change in noise levels; errors decrease as the number of observations per subject increases, given a fixed number of subjects. We next compare the estimation of Fr\'{e}chet mean $Y_{\oplus}$ (defined in Section \customref{section:2}) using our proposed size-shape decomposition method against the baseline transformation approach. We use the symbol $\widehat{\omega}_{\oplus}$ to denote Fr\'{e}chet mean estimate using the baseline transformation method. For positive functional data, $\widehat{\omega}_{\oplus}$ is defined as $\widehat{\omega}_{\oplus} = \underset{\omega \in \Omega}{\operatorname{argmin}} \ \exp{{n}^{-1} \sum_{i=1}^n\bigl(d^{2}\bigl(\log Y_i, \omega\bigr)\bigr)}$, and for monotone functional data it is  $\widehat{\omega}_{\oplus} = \underset{\omega \in \Omega}{\operatorname{argmin}} \ \bigl(\frac{\int_{0}^{t} \exp{{n}^{-1} \sum_{i=1}^n\bigl(d^{2}\bigl(\log Y_i(s), \omega(s)\bigr)\bigr)} ds} {\int_{0}^{1} \exp{{n}^{-1} \sum_{i=1}^n\bigl(d^{2}\bigl(\log Y_i(s), \omega(s)\bigr)\bigr)} ds}\bigr)$.
The estimation error for the baseline approach is thus determined by $d_{L^2}\bigl(Y_{\oplus}, \widehat{\omega}_{\oplus}\bigr)$. For the proposed decomposition method, the estimation error is given by $d_{L^2}\bigl(Y_{\oplus}, \widehat{Y}_{\oplus}\bigr)$, where $\widehat{Y}_{\oplus}$ is the Fr\'{e}chet mean estimate defined in Section \customref{section:4}. The relative root-mean-square 

\begin{figure}[ht]
    \begin{center}    
    \includegraphics[width=0.9\linewidth, height = 11cm]{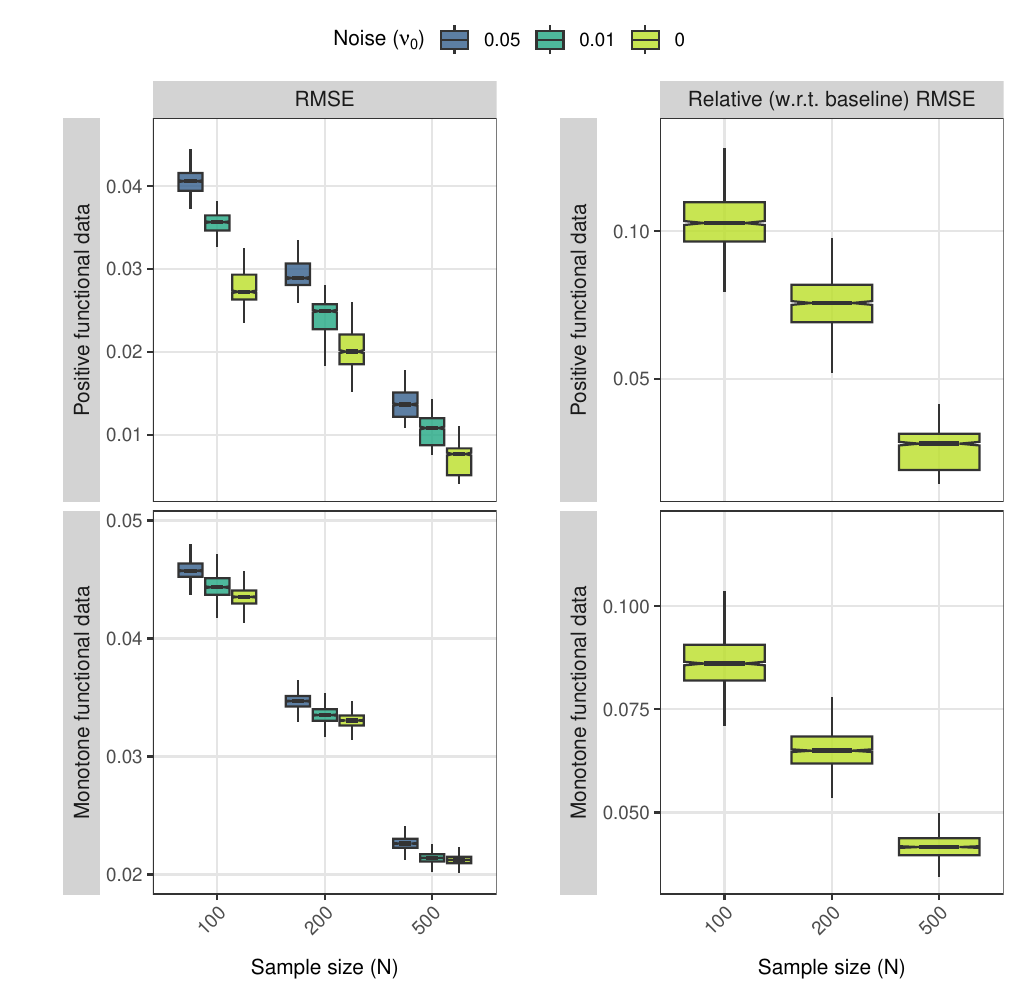}
    \caption{Simulation study for Fr\'{e}chet mean estimation based on positive and monotone functional data. (Left) RMSE for the proposed size-shape decomposition method on the number of observations $N$ available per subject and the noise variance $\nu^2_0$. (Right) Relative RMSE for the proposed method (relative to the baseline transformation approach) on the number of observations $N$ for each subject.}
    \label{fig:Sim:Posit:Mono:Frechet:Error:RATIO}
    \end{center}
\end{figure}
\noindent errors (relative RMSEs), defined as $\frac{d_{L^2}\bigl(Y_{\oplus}, \widehat{Y}_{\oplus}\bigr)}{d_{L^2}\bigl(Y_{\oplus}, \widehat{\omega}_{\oplus}\bigr)}$, are depicted in Figure \customref{fig:Sim:Posit:Mono:Frechet:Error:RATIO}. The relative RMSEs are consistently small proper fractions, demonstrating better estimation performance of our proposed method relative to the baseline transformation approach. Furthermore, an inverse relationship is observed between relative RMSE and the number of observations per subject, with the former approaching zero as the latter increases. This demonstrates that the relative performance of our method becomes increasingly better with more observations available per subject. In addition, Figure \customref{fig:Sim:Posit:Mono:Frechet:Error:RATIO}
depicts the RMSE patterns for the proposed decomposition method. The patterns, similar to those observed in Figure \customref{fig:sim-posit-mono}, further substantiate the reliability of the proposed method.

To assess the finite sample performance of global (and local) Fr\'{e}chet regression estimates, we use the following data generating mechanism for positive functional data. We generate the covariate $X \sim U(0,2)$ and the shape component (density) $f$ corresponds to a truncated normal distribution with support $[0,1]$, mean $\mu(x) = a_1 + b_1 x + \epsilon_1$, and standard deviation $\sigma(x) = a_2 + b_2 x + \epsilon_2$, where $\epsilon_1, \epsilon_2$ are independent (of other random variables too) truncated Normal $(0, \sigma_0^2)$ on support $[l_1, u_1]$ and $[l_2, u_2]$ respectively. The size component $\tau = b_3 x + \epsilon_3$, where $\epsilon_3$ is an i.i.d. copy of $\epsilon_2$. The positive trajectory is thus generated using the decomposition mapping (\customref{metric:d:on:Omega}). We set $a_1 = 0.1, b_1 = 0.3, a_2 = 0.1, b_2 = 0.1, b_3 = 0.5, \sigma_0 = 0.5, l_1 = -0.1, u_1 = 0.1, l_2 = -0.01, u_2 = 0.01$. A similar generation method is utilized for monotone functional data (see Appendix \ref{Sim:DenReg:Supp}). Figure \customref{fig:Sim:Posit:Mono:Y:GloDenReg} shows the "oracle" global Fréchet regression functions and their estimated counterparts for a simulation run, computed over a dense grid of predictor values with number of subjects $n = 500$ and $N = 500$ observations per subject. The close alignment between the oracle and estimated functions demonstrates the efficacy of our proposed approach within the Fr\'{e}chet regression framework. This visual comparison provides empirical evidence of the method's ability to accurately capture the underlying relationship between the predictor variable and functional response.

\begin{figure}[ht]
    \begin{center}
     \includegraphics[width=0.97\linewidth, height = 12cm]{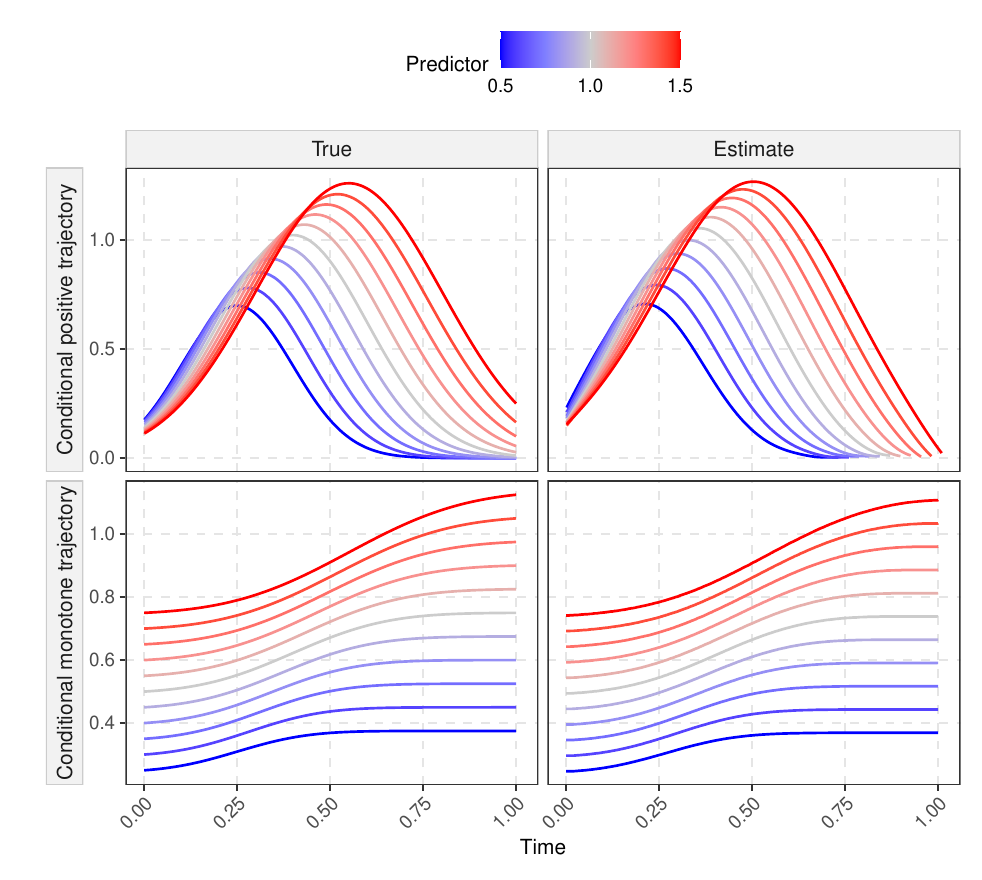}
    \caption{Conditional global Fr\'{e}chet regression functions in the simulation setting over a dense grid of predictor levels $x$, varying from $x = 0.5$ (blue) to $x = 1.5$ (red). The left panel displays the “oracle” functions and the right panel illustrates their estimated counterparts, for both positive and monotone functional data.}
    \label{fig:Sim:Posit:Mono:Y:GloDenReg}
    \end{center}
\end{figure}

\section{Data Applications}
\label{section:6}

\subsection{Medfly Activity Study}

The study of patterns of activity across the life sciences has become increasingly popular \customcitefive{fish:16}{tabac:20}{hild:23}{schr:23} {chen2023daily}. We illustrate our method for medfly activity profile data, with a lifetime activity profile available for each of $96$ female Mediterranean fruit flies (medflies). Daily locomotory activity counts were recorded for each medfly until its death (so there is no data censoring) using Monitor-LAM25 systems. The experiment involved three agar-based gel diets C10, C20 and C50 (based on sugar and yeast hydrolysate content: $10\%$, $20\%$ and $50\%$, respectively), each given to $32$ medflies. Further details on the experimental setup can be found in \customcitezero{chen2023daily}. The activity profiles are positive functional trajectories and the proposed  decomposition approach allows us to focus separately on the overall activity count (size component) and the age-varying activity pattern over life cycle (shape component). Age is measured in days and our focus is on flies that survive through a common age window $\bigl[0, 30\bigr]$ days.  
One  goal of the study was to investigate differences between diets  
and the impact of reproduction, which is quantified through the number of eggs laid by the flies 
\customcitetwo{chen2017quantifying}{chen2023daily}. 
\vspace{1mm}

\begin{figure}[ht]       
    \begin{center}
    \includegraphics[width=0.97\linewidth, height = 8cm]{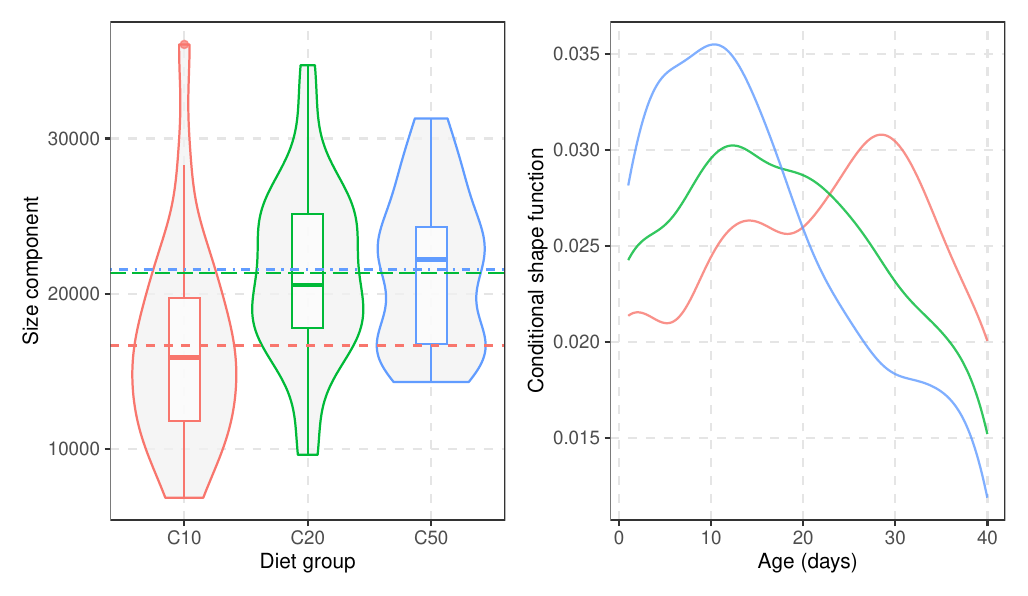}
    \caption{Medfly activity profile study with daily locomotory activity count trajectories viewed as positive functional data. (Left) Distribution of individual-level size estimates, for different diet  groups, with horizontal dashed lines representing the respective means. (Right) Conditional shape trajectories obtained for different diet groups using the decomposition-based global Fr\'{e}chet regression model.}
    \label{fig:violin:bary:medfly}
    \end{center}
\end{figure}

We first explore the variations in individual-level activity patterns in dependence on diet group. Figure \customref{fig:violin:bary:medfly} shows the distribution of subject-specific size estimates, where median size is observed to increase with higher concentrations 

\noindent of sugar and protein in the diet. Thus the overall activity counts tend to be the highest in medflies fed with C50 and lowest in C10-fed medflies. We further fit a global Fr\'{e}chet regression model to regress the shape trajectories on diet group; to mitigate boundary effects near age $30$, we use data over a slightly longer age span. The conditional shape function estimates for each diet group, depicted in Figure \customref{fig:violin:bary:medfly}, illustrate the relationship between diet and age-varying activity patterns. The diet group with the highest sugar-protein concentration, namely C50, appears to induce an initial burst of activity (until $15$ days) followed by a rapid decline, potentially indicating accelerated energy depletion. Conversely, C10-fed medflies maintain relatively higher activity levels beyond $20$ days, indicating a more sustained activity pattern or different energy utilization strategy. This

\begin{figure}[ht]
    \begin{center}
    \includegraphics[width=0.95\linewidth, height = 11cm]{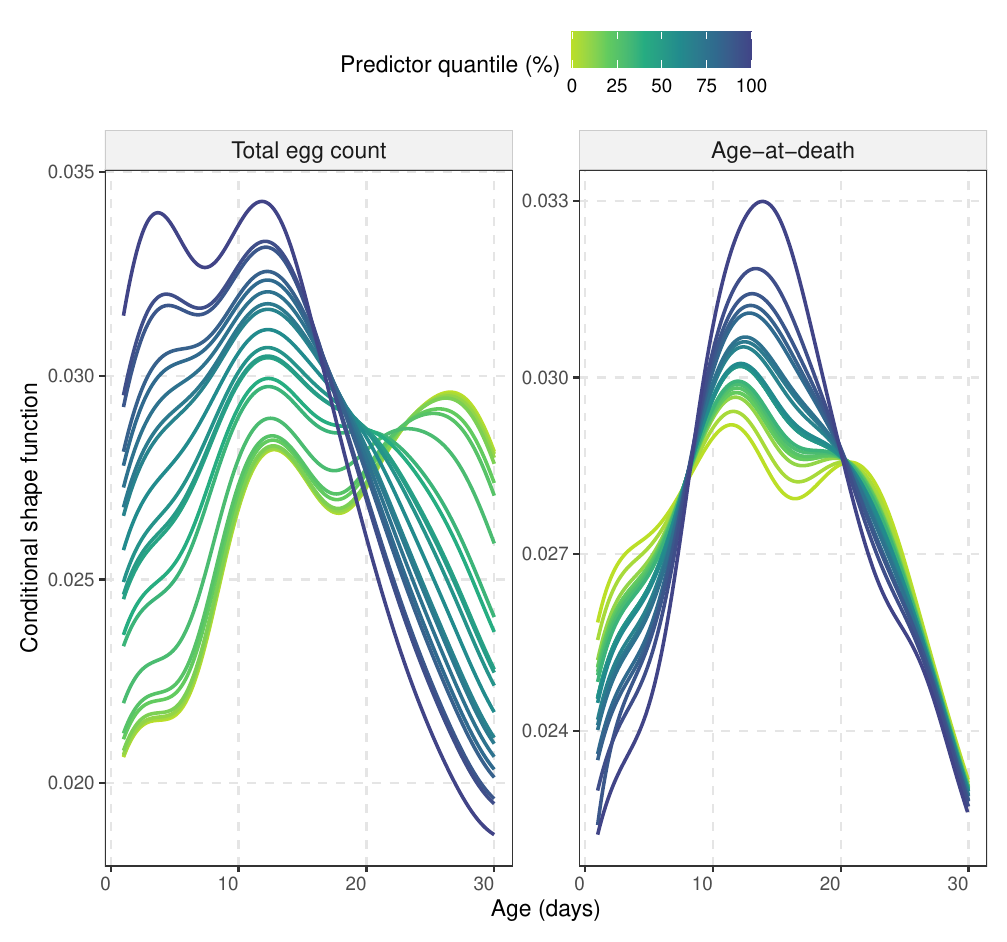}
    \caption{Medfly activity profile study with daily locomotory activity count trajectories viewed as positive functional data. Decomposition-based conditional shape functions across varying quantiles of total egg count (left) and age-at-death (right).} \label{Global:Frechet:shape:on:totegg:lifespan}
    \end{center}
\end{figure}

To investigate the relationship between reproductive magnitude on activity pattern, we fit a global Fr\'{e}chet regression model with 
(decomposition-based) shape trajectories as functional responses and total egg counts (obtained over the first $30$ days) as predictors. The conditional shape functions across varying quantiles of total egg count are presented in Figure \customref{Global:Frechet:shape:on:totegg:lifespan}, which reveals a distinct pattern. Medflies with higher total egg counts exhibit relatively higher locomo-

\noindent tory activity until around age $15$ days, followed by a lower activity thereafter. This temporal shift in activity coincides with the typical onset of egg-laying between $10$ and $15$ days, suggesting a  
link between reproductive effort and activity modulation; demonstrating how increased reproductive investment may influence energy allocation and activity patterns in  
medflies and potentially other organisms. To further study the association between activity count and lifespan, quantified as age-at-death, we fit a similar global Fr\'{e}chet regression model with age-at-death as predictor. 
The conditional shape trajectories, obtained at varying quantiles of age-at-death, are presented in Figure \customref{Global:Frechet:shape:on:totegg:lifespan}. Flies who survive longer exhibit lower initial activity levels, followed by a sharper and higher peak between ages $10$ and $20$ days, and a subsequent rapid decline. 

\begin{figure}[ht]
    \begin{center}
    \includegraphics[width=0.97\linewidth, height = 11.5cm]{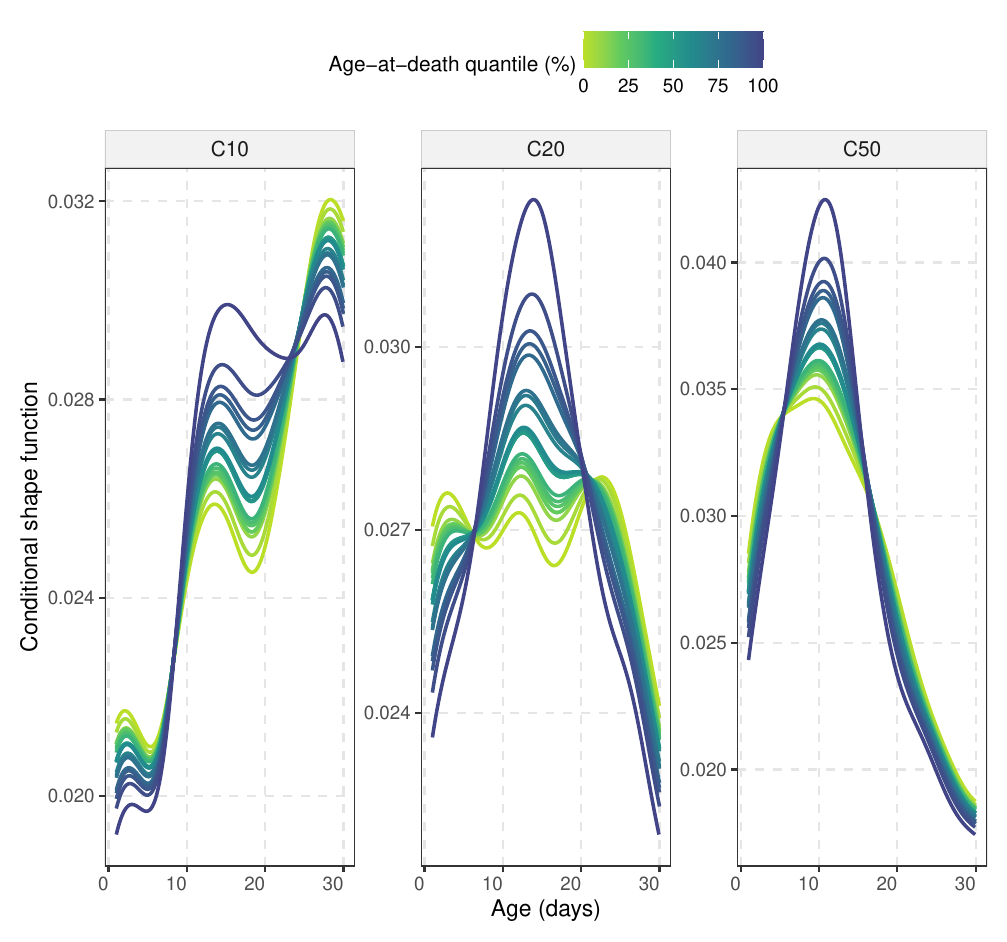}
    \caption{Medfly activity profile study with daily locomotory activity count trajectories viewed as positive functional data. Decomposition-based conditional shape functions across varying quantiles of age-at-death, and for diet groups with varying glucose-protein concentration, namely, C10 (10$\%$), C20 (20$\%$), and C50 (50$\%$).} \label{Global:Frechet:shape:on:diet:lifespan}
    \end{center}
\end{figure}

To study the effect of both diet and age-at-death on activity count, we further include diet group as an additional predictor. The conditional shape components for each diet group are illustrated in Figure \customref{Global:Frechet:shape:on:diet:lifespan}. Although the overall pattern is similar to Figure \customref{Global:Frechet:shape:on:totegg:lifespan}, there is a discernible effect of diet
group on daily activity in medflies. 

\subsection{Z\"{u}rich Longitudinal Study} 

A longitudinal study on human growth and development 
was conducted at the University Children's Hospital in Z\"{u}rich from 1954 to 1978. The growth study involved $n=232$ participants comprising $120$ males and $112$ females. Different aspects of growth, including arm length, leg length, sitting height, and standing height, were longitudinally measured at ages $1$ month, $3$ months, $6$ months, $9$ months, $1$ year, $1.5$ years and $2$ years; thereafter once every year until age $9$ years, and then every $6$ months until age $20$ years \customcitefour{bayley1964correlations}{ramsay1995comparison}{chen2012conditional}{carroll2021cross}.  The arm length trajectories 
are viewed as monotone functional data. 

We fit a decomposition-based global Fr\'{e}chet regression model with estimated height trajectories (standing or sitting) as functional responses, using the estimation approach described in Section \customref{section:4}, and sex as predictor. The conditional height trajectories illustrate distinct differences in growth patterns with respect to sex, which is not evident in the basic comparison of cross-sectional mean trajectories (See Appendix \ref{appC}, Figure \ref{Zurich:2}). 

\begin{figure}[ht]
  \begin{center}
  \includegraphics[width=0.92\linewidth, height = 10.4cm]{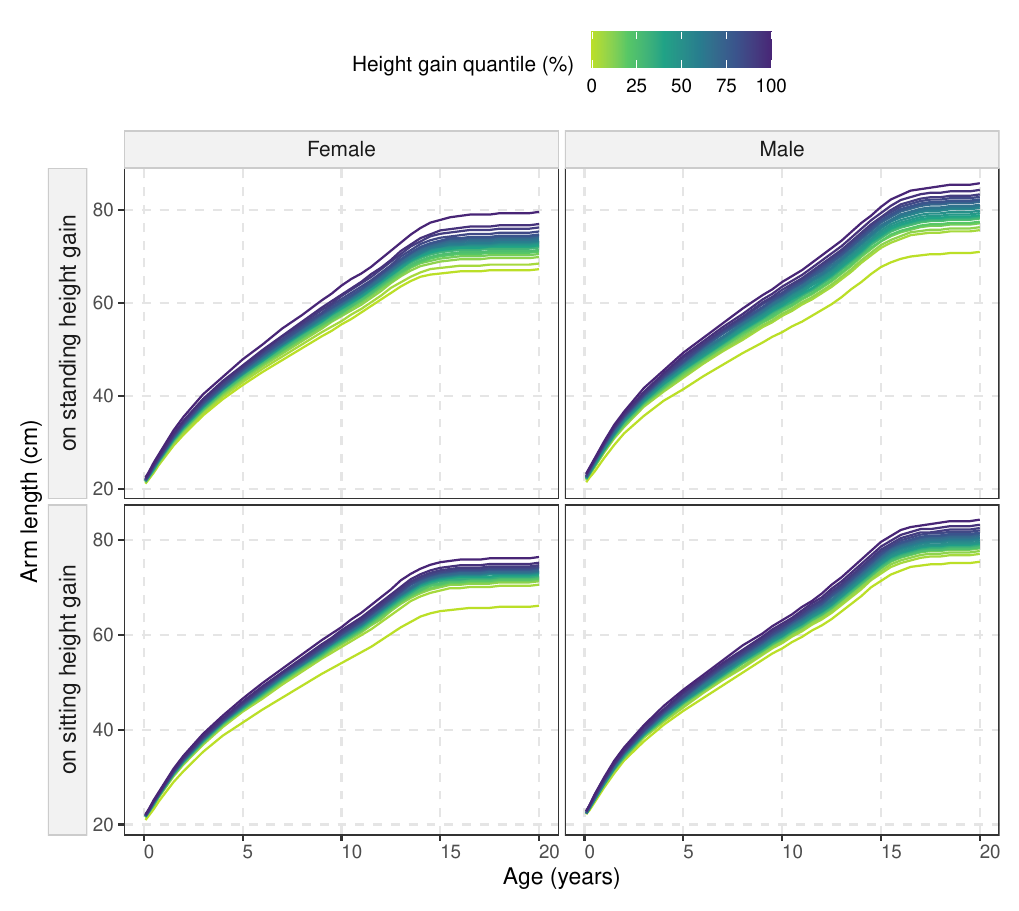} 
  \end{center}
    \caption{Z\"{u}rich longitudinal growth study with age-varying arm length trajectories viewed as monotone functional data. Decomposition-based conditional arm length trajectories across varying quantiles of standing (or sitting) height gain, for females and males.} 
    \label{Global:Frechet:Y:on:arm}
\end{figure}

To study the relationship between arm length and standing (or sitting) height gain while adjusting for sex, we regress estimated arm length trajectories on standing (or sitting) height gain in the age interval $0$ to $20$ years. Figure \customref{Global:Frechet:Y:on:arm} displays the decomposition-based conditional arm length trajectories at varying quantiles of standing (or sitting) height gain for both the sex groups. Each sex group illustrates an expected pattern, with visible difference in magnitude between males and females. 

\begin{figure}[ht]
  \begin{center}
  \includegraphics[width=0.95\linewidth, height = 9cm]{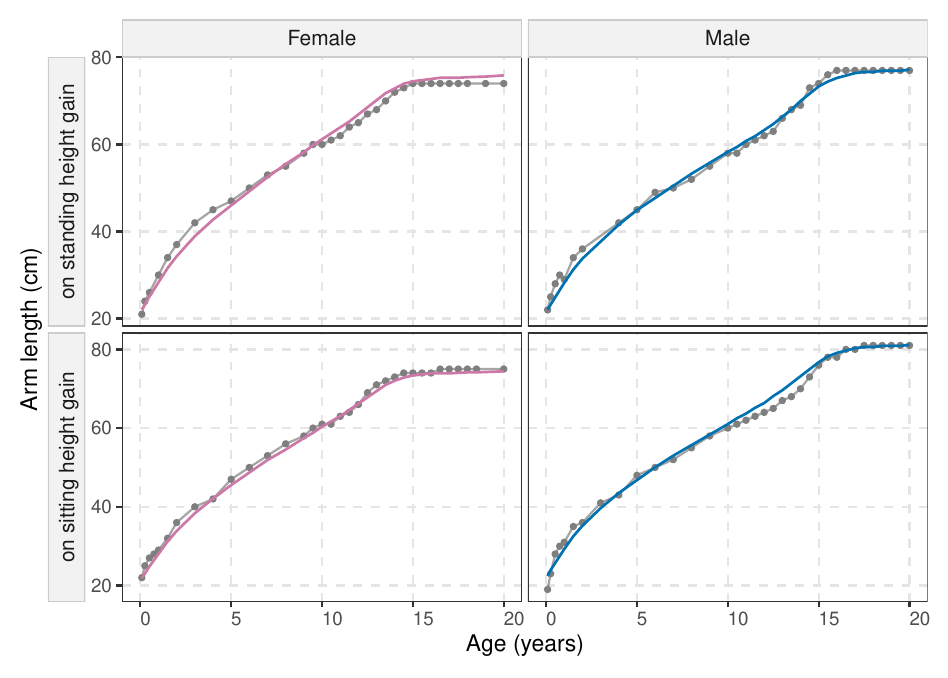} 
  \end{center}
    \caption{Z\"{u}rich longitudinal growth study with age-varying arm length trajectories viewed as monotone functional data. Observed arm length measurements (solid grey circles, with connecting lines) and predicted arm length trajectories (solid lines) for four randomly selected participants, two females (left) and two males (right). The randomly selected participants are not used in fitting the decomposition-based Fr\'{e}chet regression model.} 
    \label{pred:Y:on:arm}
\end{figure}

To assess the 
model's predictive accuracy, we randomly select male and female participants. The predicted trajectories for 
the selected individuals are obtained through a leave-one-out decomposition-based Fr\'{e}chet regression, using standing (or sitting) height gain as predictor. The predicted arm length trajectories, reported in Figure \customref{pred:Y:on:arm}, closely align with the observed arm length measurements, visually demonstrating the model's performance in capturing the association between height gain and arm length across different sex groups. 

\section{Discussion} 
\label{section:7}
In this article, we propose a decomposition-based approach for modeling shape-constrained functional data. Specifically, we decompose functional trajectories into size and shape components,  while preserving the inherent constraints, such as positivity and monotonicity, without introducing  undesirable data distortions.  Importantly, the proposed methodology does not rely on specific distributional or parametric model assumptions, hence offering enhanced applicability.  
The effectiveness of the proposed decomposition approach is further supported through rates of convergence and simulation studies. 
The proposed framework also furthers an understanding of the intrinsic geometry of shape-constrained functional data, by separating an overall size feature  from the pattern of variation over time or age. 

Several other extensions will be of interest for future research. For instance, to further ensure applicability in functional valued stochastic process \customciteone{chen2017modelling}, one may extend the proposed decomposition approach for modeling functional data subject to other shape restrictions such as convexity
or to scenarios with repeated functional data, which may also entail alternative metric choices for the decomposition components.

\begin{appendix}

\section{Additional results}
\label{appA}

\subsection{Recovery of individual trajectories}
\label{supp:theory:Thm1}
\begin{thm}
\label{Theorem:Step:Est:Yi:hat}
Under assumptions \customreftwo{ass:A1}-\customreftwo{ass:A3} and with $\gamma  \sim \bigl(\log n\big/N\bigr)^{1/3}$,
$$
    \sup_{1 \leq i \leq n}  \sup_{t \in [0, 1]} \bigl|\widehat{Y_i}(t)-Y_i(t)\bigr|=  O_p \bigl((\log n\big/N)^{1/3}\bigr),
$$
\end{thm}
where the estimation approach is discussed in Section \ref{section:4.1}.
Note that the rate of convergence $\alpha_{N,n}:=\bigl(\log n\big/N\bigr)^{1/3}$ is uniform over both the time domain and also all the subjects.

\subsection{Estimation and theory for positive functional data}
\label{supp:theory:posit}
In the local Fr\'{e}chet regression framework, we use the empirical weights from local linear regression \customciteone{fan1996local}, as adopted in  \customcitezero{petersen2019frechet}, where this concept is introduced. The empirical local weights are 
$s_{i n}(x, h)=\widehat{\sigma_0}^{-2} K_h(X_i-x)[\widehat{u}_2-\widehat{u}_1(X_i-x)]$, where $\widehat{u}_l=n^{-1} \sum_{i=1}^n K_h(X_i-x)\bigl(X_i-x\bigr)^l$ with $l \in\{0,1,2\}, \ \widehat{\sigma}_0^2=\widehat{u}_0 \widehat{u}_2-\widehat{u}_1^2$, and $K_h(\cdot)=$ $h^{-1} K(\cdot / h)$. Here the kernel $K$ satisfies Assumption \customreftwo{ass:K0} and $h:=h_n$ is a sequence of bandwidths. 
\begin{enumerate}[label = (K\arabic*), series = fregStrg, start = 0]
    \item \label{ass:K0} The kernel $K$ is a bounded 
    continuous density function, symmetric around zero. 
\end{enumerate} 
We further use estimated size $\widehat{\tau}_i$ and shape $\widehat{f}_i$ as surrogates for the latent (unobservable) decomposition components. For the regression function $\tau_{\oplus}(x) = \operatorname{argmin}_{\tau_{0} \in \Omega_{\mathcal{T}}  = [\kappa_0, \infty)} \mathbb{E}[(\tau_0 -  \tau )^2 \big\vert X=x]$, the intermediate estimate is $ \widetilde{\tau}_{\oplus}(x) = \operatorname{max} \{ \kappa_0, n^{-1} \sum_{i=1}^n s_{i n}(x, h) {\tau}_i \}$. Replacing $\tau_i$ with $\widehat{\tau}_i$, the final estimate is $ \widehat{\tau}_{\oplus}(x) = \operatorname{max} \{ \kappa_0, n^{-1} \sum_{i=1}^n s_{i n}(x, h) \widehat{\tau}_i \}$. Here $\kappa_0$ is any reasonably small positive number satisfying Assumption \customref{ass:S0}. To show the convergence of $\widehat{\tau}_{\oplus}(x)$ to $\tau_{\oplus}(x)$, we use the triangle inequality $ \bigl|\widehat{\tau}_{\oplus}(x) - \tau_{\oplus}(x)\bigr| \leq \bigl|\tau_{\oplus}(x)- \widetilde{\tau}_{\oplus}(x)\bigr|  + \bigl|\widetilde{\tau}_{\oplus}(x) - \widehat{\tau}_{\oplus}(x)\bigr|$. For the first part, $\bigl|\widehat{\tau}_{\oplus}(x)-\widetilde{\tau}_{\oplus}(x)\bigr|=\bigl| n^{-1} \sum_{i=1}^n s_{i n}(x, h)\bigl(\tau_i-\widehat{\tau}_i\bigr)\bigr| \leq n^{-1} \sum_{i=1}^n\bigl|s_{i n}(x,h)\bigr|\bigl|\widehat{\tau}_i-\tau_i\bigr|$. Therefore the convergence hinges on the consistent estimation of $\tau_i$. Using Theorem \customref{Theorem 2:uniform:tau:shape} (first part) on $\bigl|\widehat{\tau}_i-\tau_i\bigr|$, we have Proposition \customref{Proposition 3B:regg:tau:x}.
\vspace{2mm}
\begin{prop}
\label{Proposition 3B:regg:tau:x}
Suppose that the marginal density $f_X$ of $X$ satisfies $f_X (\cdot)$ $>0$ and is also twice continuously differentiable.  Under assumptions \customreftwo{ass:A1}-\customreftwo{ass:A3} and \customreftwo{ass:K0}, it holds that
\begin{equation*}
  \quad \bigl|\widehat{\tau}_{\oplus}(x)- \widetilde{\tau}_{\oplus}(x)\bigr|=O_p\bigl(N^{-1/2}\bigr). 
\end{equation*}
\end{prop}

\noindent We further make the following assumption for predictors $X$.
\begin{enumerate}[label = (L\arabic*), series = fregStrg, start = 1]
    \item \label{ass:L1} The regression function $m(x) = \mathbb{E}[\tau \big\vert X =x]$ and the marginal density $f_X(x)$ of $X$ are twice continuously differentiable in $x$. The conditional variance $\sigma^2(x) = \mathbb{E}[{(\tau - m(x))}^2 \big\vert X=x]$ is bounded and continuous.
\end{enumerate}

\noindent Under regularity conditions \customreftwo{ass:L1} and \customreftwo{ass:K0}, $\bigl|\tau_{\oplus}(x)- \widetilde{\tau}_{\oplus}(x)\bigr| = O_P\bigl(n^{-2/5}\bigr)$ with $h \sim {n^{-1 / 5}}$, using the optimal convergence rate for local linear regression \customcitetwo{fan1993local}{fan1996local}. We thus have the following result.  
\vspace{2mm}
\begin{thm}
\label{Theorem 5B:regg:tau:x}
Suppose that the marginal density $f_X$ of $X$ satisfies $f_X (\cdot)$ $>0$ and $h \sim n^{-1 / 5}$. Under assumptions \customreftwo{ass:A1}-\customreftwo{ass:A3}, \customreftwo{ass:K0}, and \customreftwo{ass:L1}, it holds that
\begin{equation*}
\bigl|\tau_{\oplus}(x) - \widehat{\tau}_{\oplus}(x)\bigr| = O_p\bigl(N^{-1/2}+n^{-2/5}\bigr).
\end{equation*}
\end{thm}

\noindent Next, we consider regressing the shape function on Euclidean predictors. Let $Q_i$ and $\widehat{Q}_i$ denote the quantile functions for the distributions associated with  $f_i$ and $\widehat{f}_i$,  respectively. Using the empirical local weights, 
\begin{equation}
\label{f:tilde:plus:x:star}
    \widetilde{f}_{\oplus}(x) =\underset{f_0 \in \Omega_\mathcal{F}}{\operatorname{argmin}}\hspace{2mm} \frac{1}{n}\sum_{i=1}^n s_{in}(x, h) d_W^2\bigl(f_i, f_0\bigr).
\end{equation}
Let $\widetilde{Q}_{\oplus}(x) : [0,1] \rightarrow [0,1]$ denote the quantile function corresponding to $\widetilde{f}_{\oplus}(x)$.
\begin{align}
\label{Q:tilde:plus:x:star}
     \widetilde{Q}_{\oplus}(\cdot, x) &=\underset{q \in \mathcal{Q}\bigl(\Omega_\mathcal{F}\bigr)}{\operatorname{argmin}} \frac{1}{n} \sum_{i=1}^n s_{in}(x, h)\bigl\|Q_i-q\bigr\|^2_{L^2([0,1])} \nonumber \\
     & = \underset{q \in \mathcal{Q}\bigl(\Omega_\mathcal{F}\bigr)}{\operatorname{argmin}} \bigl\|q-\frac{1}{n} \sum_{i=1}^n s_{i n}(x, h) Q_i\bigr\|^2_{L^2([0,1])}.  
\end{align}
The last step follows from the standard properties of the $L^2([0,1])$ inner product (Proposition 1 in \customcitezero{petersen2019frechet}). Thus $\widetilde{Q}_{\oplus}(\cdot, x)$ is the orthogonal projection of ${n}^{-1}\sum_{i=1}^n s_{i n}(x, h) Q_i$ onto the closed and convex subspace $\mathcal{Q}\bigl(\Omega_\mathcal{F}\bigr) \subset L^2([0,1])$. Hence the optimizers in (\customref{Q:tilde:plus:x:star}) and thus in (\customref{f:tilde:plus:x:star}) exist uniquely by Lemma \customref{Lemma 1}. 

\begin{lemma}
 \label{Lemma 1}
 $\mathcal{Q}\bigl(\Omega_\mathcal{F}\bigr)$ is a closed and convex subset of the Hilbert space $\mathbb{L}^2([0,1])$,
 where $\mathcal{Q}\bigl(\Omega_\mathcal{F}\bigr)$ is the space of quantile functions corresponding to the densities in $\Omega_\mathcal{F}$.
\end{lemma}
However $f_i$ (and hence $Q_i$) are not observed. Thus replacing $Q_i$ with $\widehat{Q}_i$ in $\widetilde{Q}_{\oplus}(\cdot, x)$,
\begin{equation}
\label{Q:hat:plus:x:star}
    \widehat{Q}_{\oplus}(\cdot, x) = \underset{q \in \mathcal{Q}\bigl(\Omega_\mathcal{F}\bigr)}{\operatorname{argmin}}\hspace{2mm} \bigl\|q-\frac{1}{n} \sum_{i=1}^n s_{i n}(x, h) \widehat{Q}_i\bigr\|^2_{L^2([0,1])}.  
\end{equation}

Let $\widehat{f}_{\oplus}(x)$ denote the density function corresponding to $\widehat{Q}_{\oplus}(\cdot, x)$. We now consider the convergence of $\widehat{f}_{\oplus}(x)$ to ${f}_{\oplus}(x)$, for which we first consider the triangle inequality,
$$
d_W\bigl(f_{\oplus}(x), \widehat{f}_{\oplus}(x)\bigr) \leq d_W\bigl(f_{\oplus}(x), \widetilde{f}_{\oplus}(x)\bigr)+d_W\bigl(\widetilde{f}_{\oplus}(x), \widehat{f}_{\oplus}(x)\bigr).
$$
Using properties of orthogonal projection on a closed and convex subset in Hilbert space $L^2([0,1])$, the second term $d_{W}\bigl(\widehat{f}_{\oplus}(x), \widetilde{f}_{\oplus}(x)\bigr)$ is bounded as follows.
\begin{equation}
\label{f:tilde:f:hat:plus:x:star}
    d_W\bigl(\widetilde{f}_{\oplus}(x), \widehat{f}_{\oplus}(x)\bigr) \leq \frac{1}{n} \sum_{i=1}^n\bigl|s_{i n}(x, h)\bigr|\bigl\|Q_i-\widehat{Q}_i\bigr\|_{L^2([0,1])}.
\end{equation}
Therefore the convergence hinges on consistent estimation of $Q_i$ and hence  of $f_i$. Using Theorem \customref{Theorem 2:uniform:tau:shape} (second part) on $\bigl\|Q_i-\widehat{Q}_i\bigr\|_{L^2([0,1])}$ in (\customref{f:tilde:f:hat:plus:x:star}), we have 
\vspace{2mm}
\begin{prop}
\label{Proposition 2:regg:f:x}
Suppose that the marginal density $f_X$ of $X$ satisfies $f_X (\cdot)$ $>0$ and is also twice continuously differentiable. With bin size $\gamma  \sim \bigl(\log n\big/N\bigr)^{1/3}$, under assumptions \customreftwo{ass:A1}-\customreftwo{ass:A3}, \customreftwo{ass:S0} and \customreftwo{ass:K0},
\begin{align*}
 & d_W\bigl(\widetilde{f}_{\oplus}(x), \widehat{f}_{\oplus}(x)\bigr)=O_p\bigl((\log n\big/N)^{1/6}\bigr).
\end{align*}
\end{prop}

\noindent Analogous to \customreftwo{ass:L1}, we assume \begin{enumerate}[label = (L\arabic*), series = fregStrg, start = 2]
    \item \label{ass:L2} The marginal density of $X$ and the conditional densities of $X$, given $\widetilde{Y}=y$, denoted by $f_X$ and $g_y$ respectively, exist for $\widetilde{Y} \in \Omega_{\mathcal{F}}$ and are twice continuously differentiable. The latter holds for all $y \in \Omega_{\mathcal{F}}$, with $\sup _{x, y}\bigl|g_y^{\prime \prime}(x)\bigr|<\infty$. Additionally, for any open $U \subset \Omega_{\mathcal{F}}, \int_U \mathrm{~d} F_{\widetilde{Y}  \big\vert X}(x, y)$ is continuous as a function of $x$.
\end{enumerate}

\noindent Under regularity condition \customreftwo{ass:L2}, $d_W\bigl(f_{\oplus}(x), \widetilde{f}_{\oplus}(x)\bigr) = O_P\bigl(n^{-2/5}\bigr)$ \customciteone{petersen2019frechet}. Therefore, 
\vspace{2mm}
\begin{thm}
\label{Theorem 4:regg:f:x}
Suppose that the density function satisfies $f_X(\cdot)>0$ and that  $h=h_n \sim n^{-1 / 5}$. With bin sizes $\gamma  \sim \bigl(\log n\big/N\bigr)^{1/3}$,  under assumptions \customreftwo{ass:A1}-\customreftwo{ass:A3}, \customreftwo{ass:S0}, \customreftwo{ass:K0}, and \customreftwo{ass:L2}, 
\begin{align*}
    & d_W\bigl(f_{\oplus}(x), \widehat{f}_{\oplus}(x)\bigr)=O_p\bigl((\log n\big/N)^{1/6} + n^{-2/5}\bigr).
\end{align*}
\end{thm}

\noindent By definition $Y_{\oplus}(x)=f_{\oplus}(x) \cdot \tau_{\oplus}(x)$. Therefore, a natural estimate of $Y_{\oplus}(x)$ is $\widehat{Y}_{\oplus}(x) = \widehat{f}_{\oplus}(x) \cdot \widehat{\tau}_{\oplus}(x)$. The convergence rate for $\widehat{Y}_{\oplus}(x)$ follows directly from Theorems \customref{Theorem 5B:regg:tau:x} and \customref{Theorem 4:regg:f:x}.
\vspace{2mm}
\begin{cor}
\label{Corollary 1:regg:Y:x}
Suppose that the density function satisfies $f_X(\cdot)>0$ and that $h=h_n \sim  n^{-1 / 5}$. With bin size $\gamma  \sim (\log n\big/N)^{1/3}$ and under assumptions \customreftwo{ass:A1}-\customreftwo{ass:A3}, \customreftwo{ass:S0}, \customreftwo{ass:K0} and \customreftwo{ass:L1}$-$\customreftwo{ass:L2}, 
$d \bigl(Y_{\oplus}(x), \widehat{Y}_{\oplus}(x)\bigr)=O_p\bigl((\log n\big/N)^{1/6} + n^{-2/5}\bigr).$
\end{cor}

\subsection{Estimation and theory for monotone functional data}
\label{supp:theory:mono}

The individual level estimates for size and shape components of a monotone trajectory are respectively given by,
\begin{align*}
       & \widehat{\xi}_i := \bigl(\widehat{\rho}_i, \widehat{\lambda}_i\bigr), \text{ with }  \widehat{\rho}_i = \widehat{Y}_i(1) -  \widehat{Y}_i(0) , \ \widehat{\lambda}_i = \widehat{Y}_i(0)  \ \ \text{and} \nonumber\\
    &  \widehat{F}_i(t):= \frac{\widehat{Y}_i(t) - \widehat{\lambda}_i}{\widehat{\rho}_i} 
    , \ \ \text{for all } t \in [0,1].
\end{align*}

\begin{thm}
\label{Theorem 2B:uniform:tau:mono}
Under assumptions \customreftwo{ass:A1}-\customreftwo{ass:A3} and with bin size $\gamma  \sim \bigl(\log n\big/N\bigr)^{1/3}$, the size component estimate $\widehat{\xi}_i$ satisfies
\begin{align*}    
&\sup_{1 \leq i \leq n} \bigl|\widehat{\rho}_i-\rho_i\bigr|=O_p\bigl((\log n\big/N)^{1/3}\bigr),  \
\sup_{1 \leq i \leq n} \bigl|\widehat{\lambda}_i-\lambda_i\bigr| = O_p\bigl((\log n\big/N)^{1/3}\bigr), \\
& \text{and therefore,} \ \sup_{1 \leq i \leq n} d_E \bigl(\widehat{\xi}_i,\xi_i\bigr) = O_p\bigl((\log n\big/N)^{1/3}\bigr).
    \end{align*}
\end{thm}

\noindent For the shape component $F$, we additionally assume

\begin{enumerate}[label = (S\arabic*), series = fregStrg, start = 1]
    \item \label{ass:S1} There exists a scalar $\kappa_2 > 0$ such that all realizations of the underlying stochastic process $Y(t)$ satisfy $\bigl| Y_i(1) - Y_i(0)\bigr| \geq \kappa_2$.
\end{enumerate}
Assumption~\customreftwo{ass:S1} ensures that the range component $\rho$ is bounded away from 0, since $\rho=0$ corresponds to a constant trajectory and is not of much interest. Thus \customreftwo{ass:S1} in turn ensures that $F(t) = (Y(t)-Y(0))\big/\rho$ is well defined. 
\vspace{2mm}
\begin{thm}
\label{Theorem 3:uniform:f:mono}
 Under assumptions \customreftwo{ass:A1}-\customreftwo{ass:A3} and \customreftwo{ass:S1}, and with bin size $\gamma  \sim \bigl(\log n\big/N\bigr)^{1/3}$,  the shape component estimate $\widehat{F}_i(t)$ satisfies
\begin{align*}
    \sup_{1 \leq i \leq n}  \sup_{t \in [0, 1]} \bigl|\widehat{F}_i(t)-F_i(t)\bigr| &=  O_p\bigl((\log n\big/N)^{1/3}\bigr) \ \text{and} 
    \\ 
    \sup_{1 \leq i \leq n}  d_{\mathbb{L}^2([0,1])} \bigl(\widehat{Q}_i,Q_i\bigr) &=O_p\bigl((\log n\big/N)^{1/6}\bigr),
\end{align*}
where $\widehat{Q}_i$ and ${Q}_i$ are quantile functions associated with the CDFs $\widehat{F}_i$ and $F_i$, respectively.
\end{thm}

Next, we address estimation of the Fr\'{e}chet means for the decomposition components. For the size components, we propose $\widehat{\xi}_{\oplus} = \bigl(\widehat{\rho}_{\oplus},\widehat{\lambda}_{\oplus}\bigr)$, with $\widehat{\rho}_{\oplus} = n^{-1} \sum_{i=1}^{n} \widehat{\rho}_i $ and $\widehat{\lambda}_{\oplus} =  n^{-1} \sum_{i=1}^{n} \widehat{\lambda}_i $. For $F_{\oplus}$, the estimator $\widehat{F}_{\oplus}$ is defined as the distribution function corresponding to $\widehat{Q}_{\oplus} := \underset{q \in {\mathcal{Q}}(\Omega_{\mathcal{F}})}{\operatorname{argmin} } \ \bigl\| q - n^{-1} \sum_{i=1}^{n} \widehat{Q}_i \bigr\|_{L^2([0,1])}^2$. Consequently, $\widehat{Y}_{\oplus}=\widehat{\lambda}_{\oplus}+\widehat{\rho}_{\oplus} \widehat{F}_{\oplus}$.
\vspace{1mm}
\begin{cor}
    \label{Corollary:Tau:f:Y:plus:mono}
     Under assumptions \customreftwo{ass:A1}-\customreftwo{ass:A3} and bin size $\gamma \sim (\log n\big/N)^{1/3}$, $|\widehat{\rho}_{\oplus} - \rho_{\oplus}|= O_p\bigl((\log n\big/N)^{1/3} + n^{-1/2} \bigr), \  |\widehat{\lambda}_{\oplus} - \lambda_{\oplus}|= O_p\bigl((\log n\big/N)^{1/3} + n^{-1/2} \bigr), $ \\ $
    \text{and therefore, } d_E(\widehat{\xi}_{\oplus} - \xi_{\oplus})= O_p\bigl((\log n\big/N)^{1/3} + n^{-1/2} \bigr).
     $
     \\ Additionally with Assumption \customreftwo{ass:S1}, $ d_{W}(F_{\oplus}, \widehat{F}_{\oplus}) = O_p\bigl((\log n\big/N)^{1/6} + n^{-1/2}\bigr).$ Combining both,  $d\bigl(Y_{\oplus}, \widehat{Y}_{\oplus}\bigr) = O_p\bigl((\log n\big/N)^{1/6} + n^{-1/2}\bigr)$.
\end{cor}
\vspace{1mm}
Subsequently in local Fr\'{e}chet regression, the estimators for $\rho_{\oplus}(x) = \\ {\operatorname{argmin}}_{\rho_{0} \in \Omega_{\mathcal{T}}  = [\kappa_2, \infty)} \mathbb{E}\bigl[(\rho_0 -\rho)^2 \big\vert X=x\bigr], \ \lambda_{\oplus}(x) = \mathbb{E}\bigl[\lambda \big\vert X=x\bigr]$, and $\xi_\oplus(x) = \bigl({\rho}_{\oplus}(x), {\lambda}_{\oplus}(x)\bigr)$ are respectively $\widehat{\rho}_{\oplus}(x) = \max \{\kappa_2, n^{-1} \sum_{i=1}^n s_{i n}(x, h) \widehat{\rho}_i \}, \ \widehat{\lambda}_{\oplus}(x) \\ = n^{-1} \sum_{i=1}^n s_{i n}(x, h) \widehat{\lambda}_i$, and $\widehat{\xi}_{\oplus}(x) = \bigl(\widehat{\rho}_{\oplus}(x), \widehat{\lambda}_{\oplus}(x) \bigr)$. Here $\kappa_2$ is any reasonably small positive number satisfying Assumption \customref{ass:S1}. The distribution function $\widehat{F}_{\oplus}(x)$ corresponding to $\widehat{Q}_{\oplus}(\cdot, x)$, defined in (\customref{Q:hat:plus:x:star}), is proposed as an estimator for $F_{\oplus}(x)$. Following the definition of $Y_{\oplus}(x)$, we define the estimator $\widehat{Y}_{\oplus}(x) = \widehat{\lambda}_{\oplus}(x) + \widehat{\rho}_{\oplus}(x) \cdot \widehat{F}_{\oplus}(x)$. We thus have the following convergence result. 
\vspace{2mm}
\begin{thm}
\label{Theorem 5B:regg:tau:F:x:mono}
Suppose that the marginal density $f_X$ of $X$ satisfies $f_X (\cdot)$ $>0$ and that $h \sim n^{-1 / 5}$. With bin size $\gamma  \sim (\log n\big/N)^{1/3}$, and under assumptions \customreftwo{ass:A1}-\customreftwo{ass:A3},  \customreftwo{ass:K0}, \customreftwo{ass:L1},
\begin{align*}
&|\rho_{\oplus}(x) - \widehat{\rho}_{\oplus}(x)| = O_p\bigl((\log n\big/N)^{1/3} + n^{-2/5}\bigr),  \ |\lambda_{\oplus}(x) - \widehat{\lambda}_{\oplus}(x)| = O_p\bigl((\log n\big/N)^{1/3} \bigr. \nonumber \\ &\bigl. + \ n^{-2/5}\bigr),  \ \text{and} \ d_E (\xi_{\oplus}(x) , \widehat{\xi}_{\oplus}(x)) = O_p\bigl((\log n\big/N)^{1/3} + n^{-2/5}\bigr).
\end{align*}
\text{Additionally, under assumptions \customreftwo{ass:S1} and \customreftwo{ass:L2} (instead of \customreftwo{ass:L1}),}
$$d_W\bigl(F_{\oplus}(x), \widehat{F}_{\oplus}(x)\bigr) =O_p\bigl((\log n\big/N)^{1/6} + n^{-2/5}\bigr).$$
\text{Under all the above assumptions, we have} $$d \bigl(Y_{\oplus}(x), \widehat{Y}_{\oplus}(x)\bigr)=O_p\bigl((\log n\big/N)^{1/6} + n^{-2/5}\bigr).$$
\end{thm} 
  
\noindent Again, to enable the inclusion of categorical predictors in the model, we consider estimating the global Fr\'{e}chet regression function for the decomposition components. Let $\widehat{\rho}_{G_\oplus}(x) = \max \{ \kappa_2,\frac{1}{n} \sum_{i=1}^n s_{i n}(x) \widehat{\rho}_i \}$, $\widehat{\lambda}_{G_\oplus}(x) = \frac{1}{n} \sum_{i=1}^n s_{i n}(x) \widehat{\lambda}_i $, and $\widehat{\xi}_{G_\oplus}(x)=(\widehat{\rho}_{G_\oplus}(x),\widehat{\lambda}_{G_\oplus}(x))$ be the proposed estimators for 
$\rho_{G_\oplus}(x)$, $\lambda_{G_\oplus}(x)$, and $\xi_{G_\oplus}(x)=(\rho_{G_\oplus}(x),\lambda_{G_\oplus}(x))$, respectively. The proposed estimator for ${F}_{G_\oplus}(x)$ is the distribution function $\widehat{F}_{G_\oplus}(x)$, corresponding to $\widehat{Q}_{G_\oplus}(\cdot, x)$ defined in Section \customref{section:3}. Consequently, we estimate $Y_{G_\oplus}(x)$ using the estimator $\widehat{Y}_{G_\oplus}(x) = \widehat{\lambda}_{G_\oplus}(x) + \widehat{\rho}_{G_\oplus}(x) \cdot \widehat{F}_{G_\oplus}(x)$. The theorem below gives the relevant rate of convergence.

\begin{thm}
\label{Theorem 8B:Global:regg:tau:Gplus:x:mono}
Under assumptions \customreftwo{ass:A1}-\customreftwo{ass:A3} and with bin size $\gamma  \sim (\log n\big/N)^{1/3}$, 
\begin{align*}
& |\rho_{G_{\oplus}}(x) - \widehat{\rho}_{G_{\oplus}}(x)| = O_p\bigl((\log n\big/N)^{1/3} + n^{-1/2}\bigr), \ |\lambda_{G_{\oplus}}(x) - \widehat{\lambda}_{G_{\oplus}}(x)| = O_p\bigl((\log n \big/   \bigr. \nonumber \\
& \bigl.  N)^{1/3} + n^{-1/2}\bigr), \ \text{and } d_E (\xi_{G_{\oplus}}(x) , \widehat{\xi}_{G_{\oplus}}(x)) = O_p\bigl((\log n\big/N)^{1/3} +n^{-1/2}\bigr).\\ &  \text{Additionally under Assumption \customreftwo{ass:S1},} \ d_W(F_{G_{\oplus}}(x),  \widehat{F}_{G_{\oplus}}(x))=O_p\bigl((\log n\big/N)^{1/6} + \bigr. \nonumber \\
& \bigl. n^{-1/2}\bigr) \ \text{and} \ d (Y_{G_{\oplus}}(x), \widehat{Y}_{G_{\oplus}}(x))=O_p\bigl((\log n\big/N)^{1/6} + n^{-1/2}\bigr).
\end{align*}
\end{thm}

\subsection[]{Extension: Non-equidistant time grid with varying $N_i$} 
\label{sp: section 13}

Instead of equidistant time grid, here we consider the setup involving non-equidistant time grid with $N_i$ observations for the $i$th subject. The estimate $\widehat{Y}_i(t)$ of $Y_i(t)$ retains the form, 
\begin{align*}
\widehat{Y}_i(t) =  \frac{\sum_{j=1}^{N_i} \ Z_{ij} \mathds{1}_{\{t_{ij} \in B_\ell\}}}{\sum_{j=1}^{N_i} \ \mathds{1}_{\{t_{ij} \in B_\ell\}}}, \: \text{for} \ t \in B_\ell \ \text{and} \ \ell = 1, \dots, L,
\end{align*}
where $L=\gamma^{-1}$, and $\gamma$ denotes the common bin width. Let $\Delta_i := \underset{1 \leq \ell \leq L}{\max} \ | t_{i\ell} - t_{i(\ell-1)} |$
denote the maximum spacing in the grid for the $i$th subject. Therefore, $M_{i,\ell} \geq \gamma / \Delta_i$. Recall the notation $M_{i,\ell}$ denoting the number of observations in the $\ell$th bin for the $i$th subject. We assume the following bound on the maximum grid spacing in order to establish a uniform convergence rate.

\begin{enumerate}
[label = (A\arabic*), series = fregStrg, start = 4]
    \item \label{ass:A4} 
There exists a positive sequence $\Delta_0:=\Delta_0(n)$ such that $\underset{1 \leq i \leq n}{\max} \Delta_i \leq \Delta_0$, $\Delta_0 \rightarrow 0$ and $\Delta_0 \log n  \rightarrow 0$ as $n \rightarrow \infty$.     
\end{enumerate} 

\noindent By Assumption \customreftwo{ass:A4}, it follows that $\underset{1 \leq i \leq n}{\inf} \ \underset{\ell=1, \dots, L}{\inf}M_{i,\ell} \geq \Delta^{-1}_0 \gamma$. Following similar logic as outlined in the proof of Theorem~\customreftwo{Theorem:Step:Est:Yi:hat}, the optimal bin width $\gamma$ is obtained as a solution of the equation $\gamma^2 = \log (n\gamma^{-1}) \big/ \Delta_0^{-1} \gamma$. The choice $\gamma = {(\Delta_0 \log n)}^{1/3}$ results in approximate equality between the two sides. Thus, we have the following convergence result.
\begin{thm}
\label{Theorem:Step:Est:Yi:hat:Non:Equi}
Under assumptions \customreftwo{ass:A1}, \customreftwo{ass:A2}, \customreftwo{ass:A4} and bin size $\gamma  \sim \bigl(\Delta_0 \log n\bigr)^{1/3}$, the estimate $\widehat{Y}_i(t)$ defined in equation (\customref{Est:Yi:Step}) of the main paper satisfies
\begin{align*}
    \sup_{1 \leq i \leq n} \underset{t \in [0, 1]}{\sup} \ \bigl|\widehat{Y}_i(t)-Y_i(t)\bigr|=  O_p\bigl((\Delta_0 \log n)^{1/3}\bigr).
\end{align*}
\end{thm}
\noindent The convergence rate $\theta_{\Delta_0,n}:=\bigl(\Delta_0 \log n\bigr)^{1/3}$ is uniform over all the subjects as well as the time domain. 
For positive functional data with non-equidistant time grid, the estimator $\frac{1}{N_i} \sum_{j=1}^{N_i} Z_{ij}$ for size component $\tau_i=\int_0^{1} Y_i(t) d t$ is replaced with $\widetilde{\tau}_i=\int_0^{1} \widehat{Y}_i(t) d t$. The corresponding rate of convergence is then established following similar steps as in the proof of Theorem \customref{Theorem 2:uniform:tau:shape}, Eq. 33. For non-equidistant time grid, the rate $\theta_{\Delta_0,n}$ will be carried forward instead of $\alpha_{N,n}$ in all the subsequent results, and will mirror the same logical progression.

\section{Additional plots}
\label{appC}

\subsection{Simulation for decomposition-based Fr\'{e}chet regression}
\label{Sim:DenReg:Supp}
\begin{figure}[H]
    \begin{center}
     \includegraphics[width=0.95\linewidth, height = 10cm]{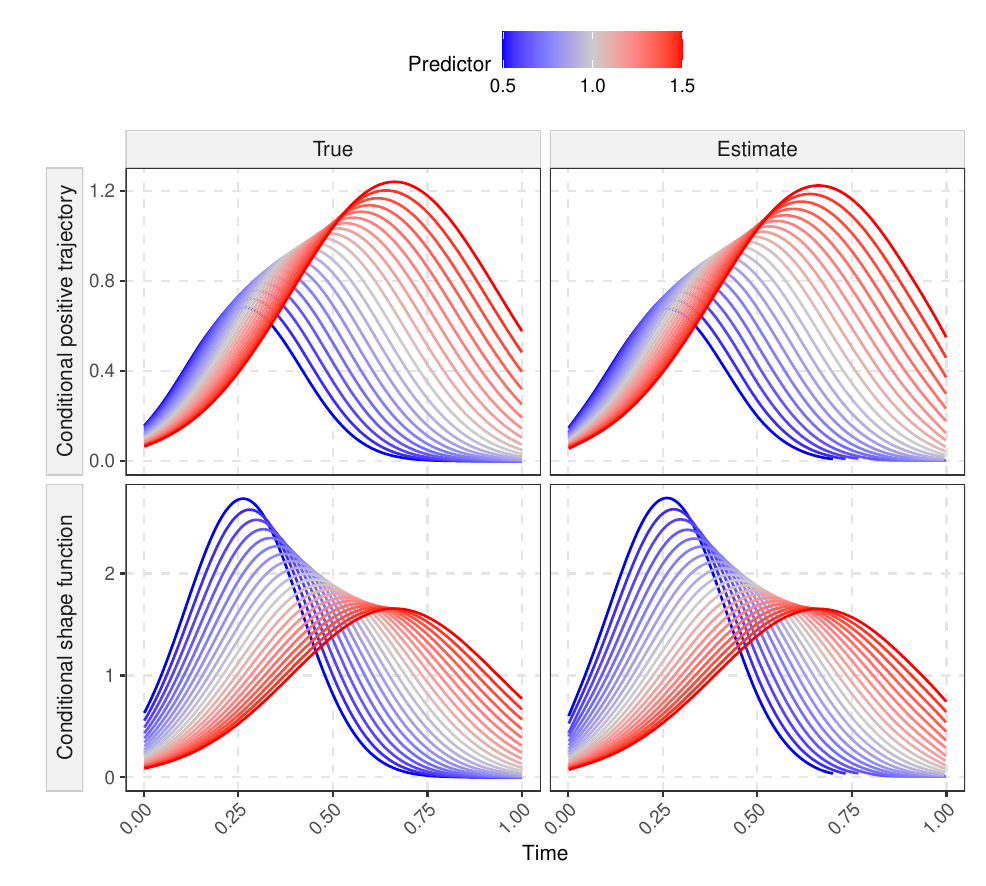}
    \caption{Conditional local Fr\'{e}chet regression functions in the simulation setting over a dense grid of predictor levels $x$, varying from $x = 0.5$ (blue) to  $x = 1.5$ (red). The left panel displays the “oracle” functions and the right panel illustrates their estimated counterparts, for positive trajectories and their shape components.}
    \label{fig:Sim:Posit:LocDenReg}
    \end{center}
\end{figure}

We assess the finite sample performance of local Fr\'{e}chet regression estimates for both the positive trajectories and their associated shape components. We use the following data generating mechanism. We generate the continuous covariate $X$ from truncated normal $(\mu_X, \sigma_X)$ with support $[l_X, u_X]$. The shape component (density) $f$ corresponds to a truncated normal distribution with support $[0,1]$, mean $\mu(x) = a_1 + b_1 x + c_1 x^2 + \epsilon_1$, and standard deviation $\sigma(x) = a_2 + b_2 x + c_2 x^2 + \epsilon_2$, where $\epsilon_1, \epsilon_2$ are independent (of other random variables too) truncated Normal $(0, \sigma_0)$ on support $[l_1, u_1]$ and $[l_2, u_2]$ respectively. The size component is  $\tau = b_3 x + \epsilon_3$, where $\epsilon_3$ is an i.i.d. copy of $\epsilon_2$. The positive trajectory is generated using the decomposition mapping from equation (\customref{metric:d:on:Omega}) of Section \customref{section:2}. We choose $a_1 = 0.1, b_1 = 0.3, c_1 = 0.05, a_2 = 0.1, b_2 = 0.1, c_2 = 0.01, b_3 = 0.5, \sigma_0 = 0.5, \mu_X = 1, \sigma_X = 0.5, l_X = 0, u_X = 2, l_1 = -0.1, u_1 = 0.1, l_2 = -0.01, u_2 = 0.01$. Figure \customref{fig:Sim:Posit:LocDenReg} shows the “oracle” local Fr\'{e}chet regression function and their estimated counterparts for one simulation run over a dense grid of predictor values, for $n = 500$ and $N = 500$.
\begin{figure}[H]
    \begin{center}
     \includegraphics[width=0.95\linewidth, height = 10cm]{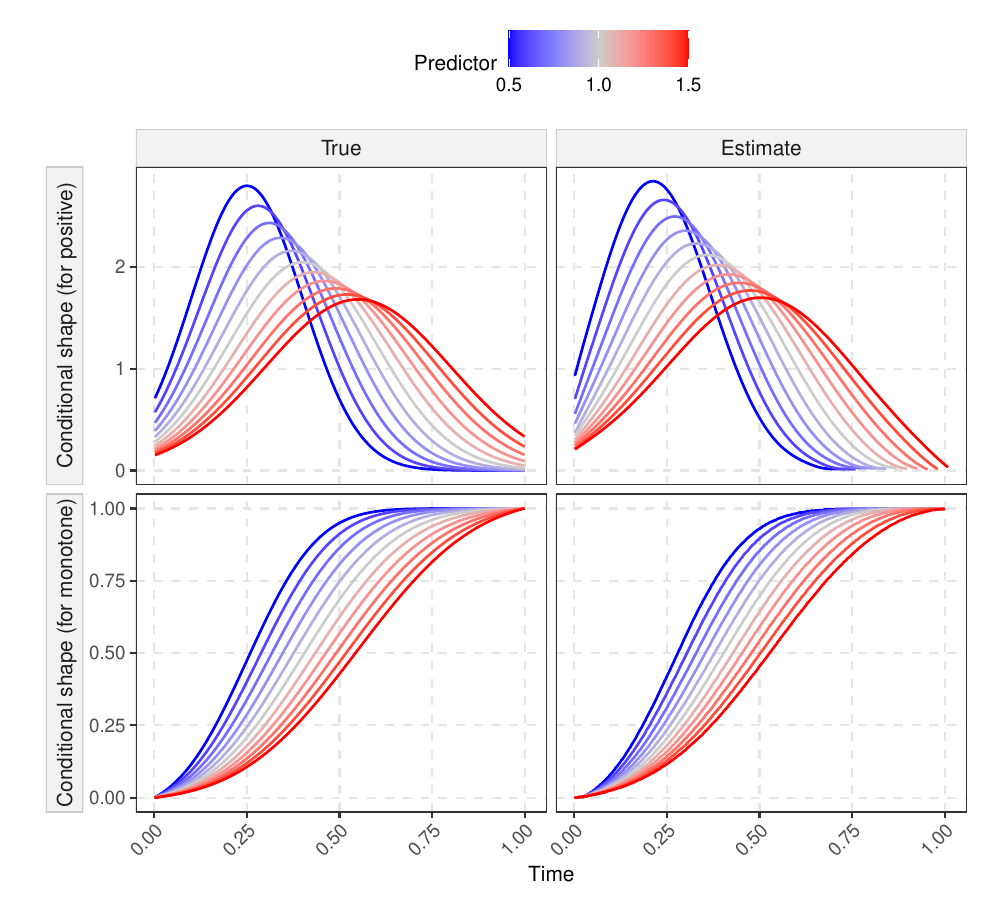}
    \caption{Conditional global Fr\'{e}chet regression functions for shape trajectories over a dense grid of predictor levels $x$, varying from $x = 0.5$ (blue) to  $x = 1.5$ (red). The left panel displays the “oracle” functions and the right panel illustrates their estimated counterparts, for both positive and monotone functional data.}
    \label{fig:Sim:Posit:Mono:shape:GloDenReg}
    \end{center}
\end{figure}

For monotone functional data, we use a similar simulation setup as in the case of positive functional data. The shape component $F$ is the cumulative distribution function of a truncated normal distribution with support $[0,1]$, mean $\mu(x) = a_1 + b_1 x + c_1 x^2 + \epsilon_1$, and standard deviation $\sigma(x) = a_2 + b_2 x + c_2 x^2 + \epsilon_2$. The notations and parameter values are consistent with the previous setup. The size components are $\lambda = b_3 x + \epsilon_3$ and $\rho = b_4 x + \epsilon_4$ with $b_3 = 0.5, b_4 = 0.25$, where $\epsilon_3, \epsilon_4$ are i.i.d. copies of $\epsilon_2$. The monotone trajectory is then generated using the decomposition mapping from equation (\customref{metric:d:on:Omega:mono}) of Section \customref{section:2}.

\begin{figure}[H]
    \begin{center}
     \includegraphics[width=0.95\linewidth, height = 10cm]{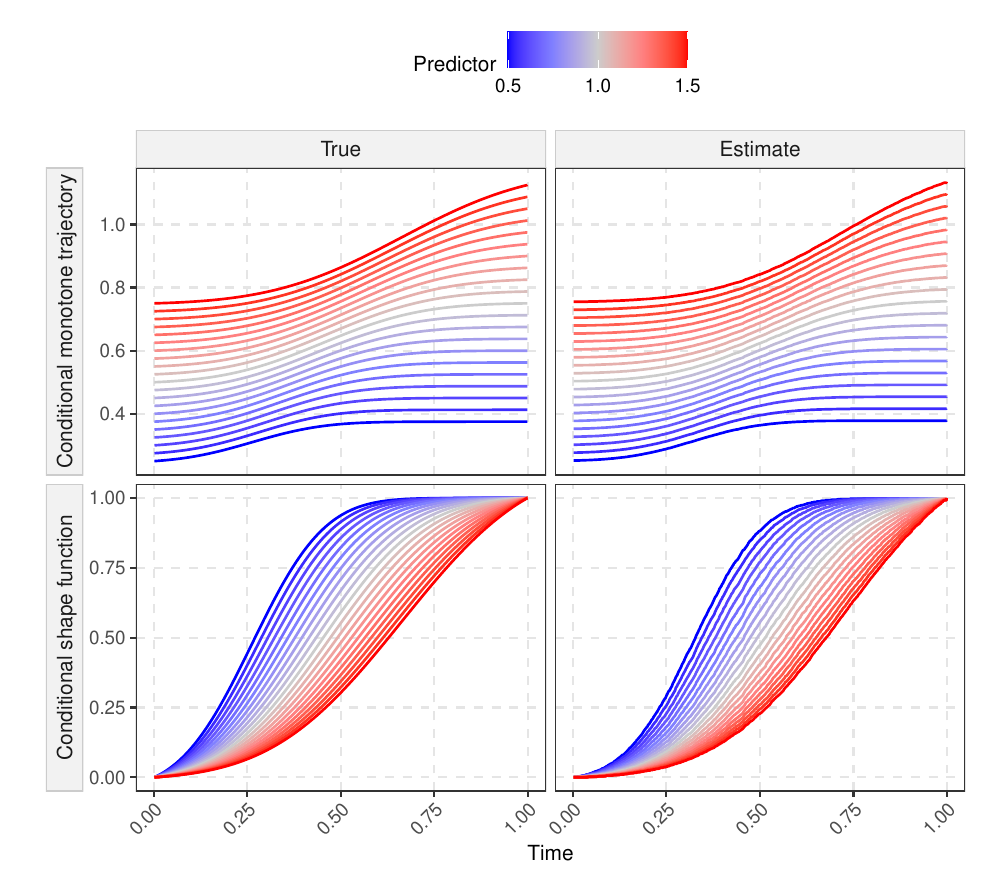}
    \caption{Conditional local Fr\'{e}chet regression functions in the simulation setting over a dense grid of predictor levels $x$, varying from $x = 0.5$ (blue) to  $x = 1.5$ (red). The left panel displays the “oracle” functions and the right panel illustrates their estimated counterparts, for monotone trajectories and their shape components.}
    \label{fig:Sim:Mono:LocDenReg}
    \end{center}
\end{figure}

\subsection{Medfly activity profile data application}
\begin{figure}[H]
    \begin{center}
    \includegraphics[width=0.95\linewidth, height = 9.6cm]{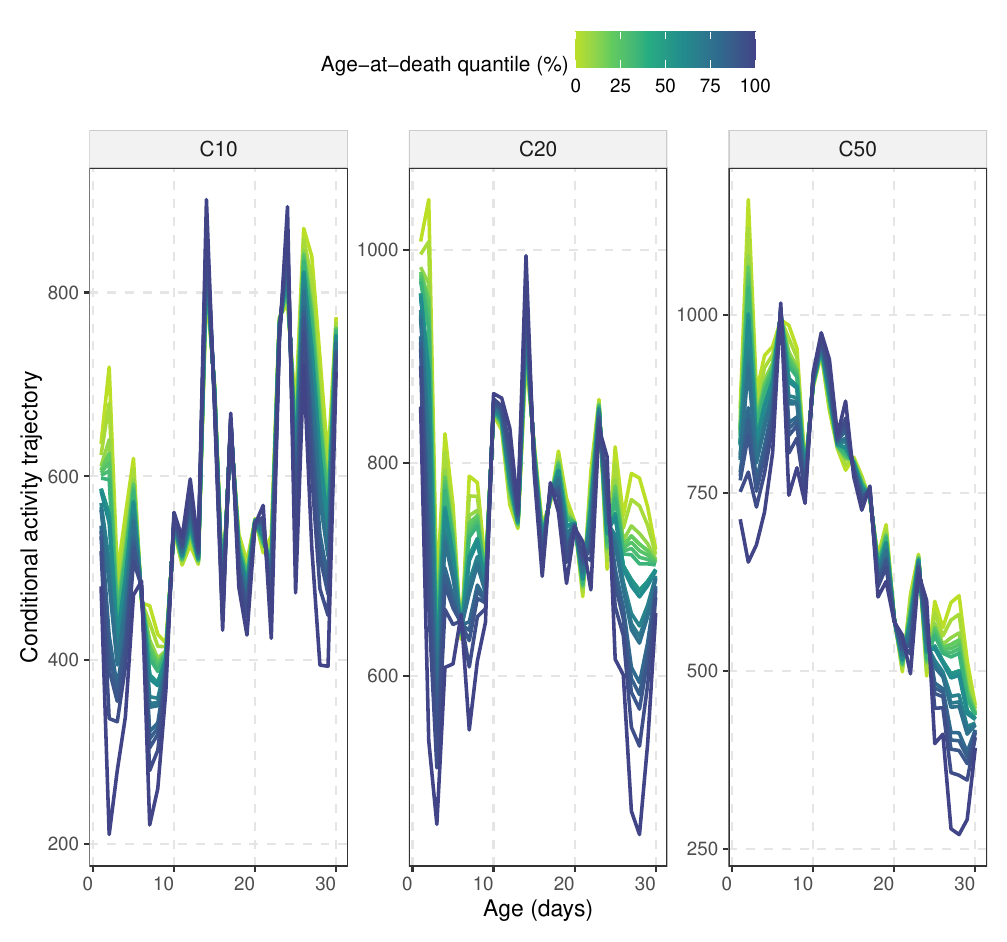}
    \includegraphics[width=0.95\linewidth, height = 9.6cm]{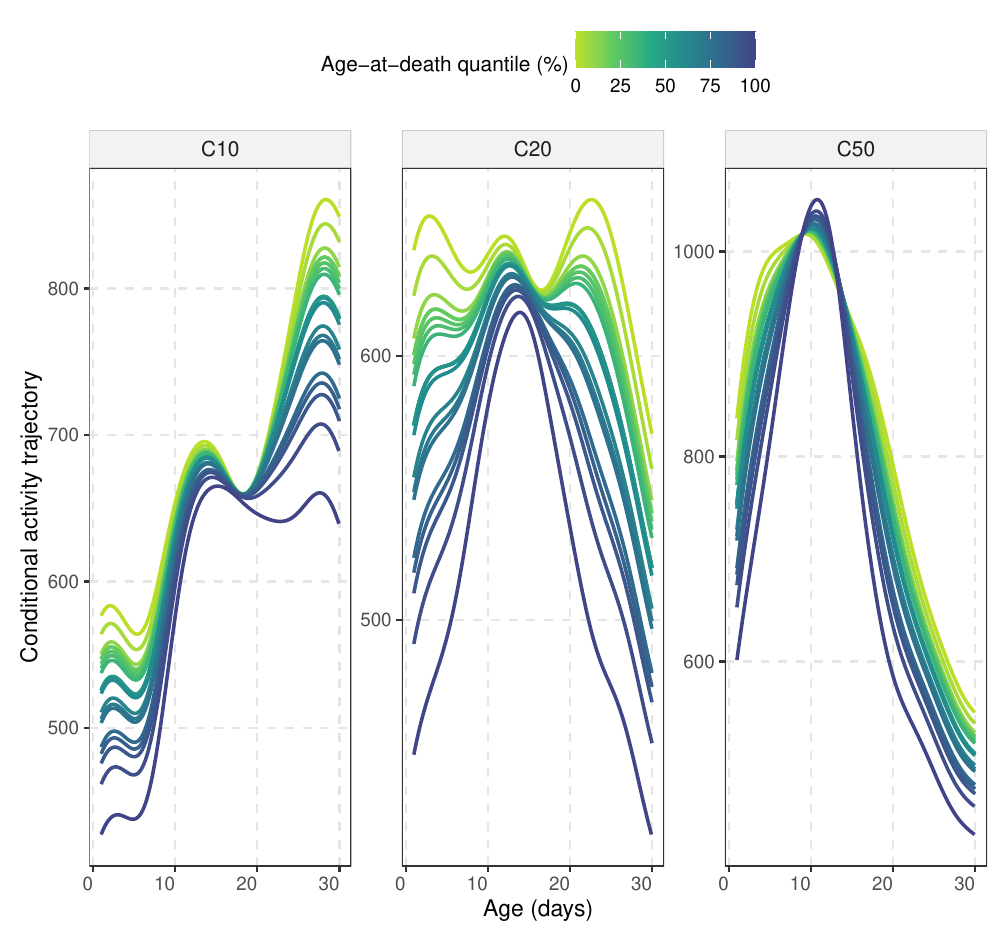}
    \caption{ Conditional activity trajectories across varying quantiles of age at death, based on regular (top panel) and decomposition-based (bottom panel) global Fr\'{e}chet regression model.} \label{Global:Frechet:Y:on:lifespan:nut}
    \end{center}
\end{figure}

To explore the effect of survival duration and diet on locomotory activity patterns, we fit a decomposition-based global Fr\'{e}chet regression model with (estimated) activity trajectories as functional response. We use the decomposition metric $d$ defined in equation (\customref{metric:d:on:Omega}) of Section \customref{section:2}. The resulting conditional activity trajectories, obtained at five equidistant sample quantiles of age at death, are illustrated in Figure \customref{Global:Frechet:Y:on:lifespan:nut}. Within each diet group, flies with longer lifespan tend to demonstrate relatively lower daily activity counts. The trajectories overlap at some point between $10$ to $20$ days. This minimal divergence is likely associated with egg-laying activity of the flies, since the flies start laying eggs typically after $10$ days. However, a conventional global Fr\'{e}chet regression model with functional response fails to elucidate the pattern.

\subsection{Z\"{u}rich growth study data application}
\begin{figure}[H]
    \centering
\includegraphics[width=0.8\linewidth,height=7cm]{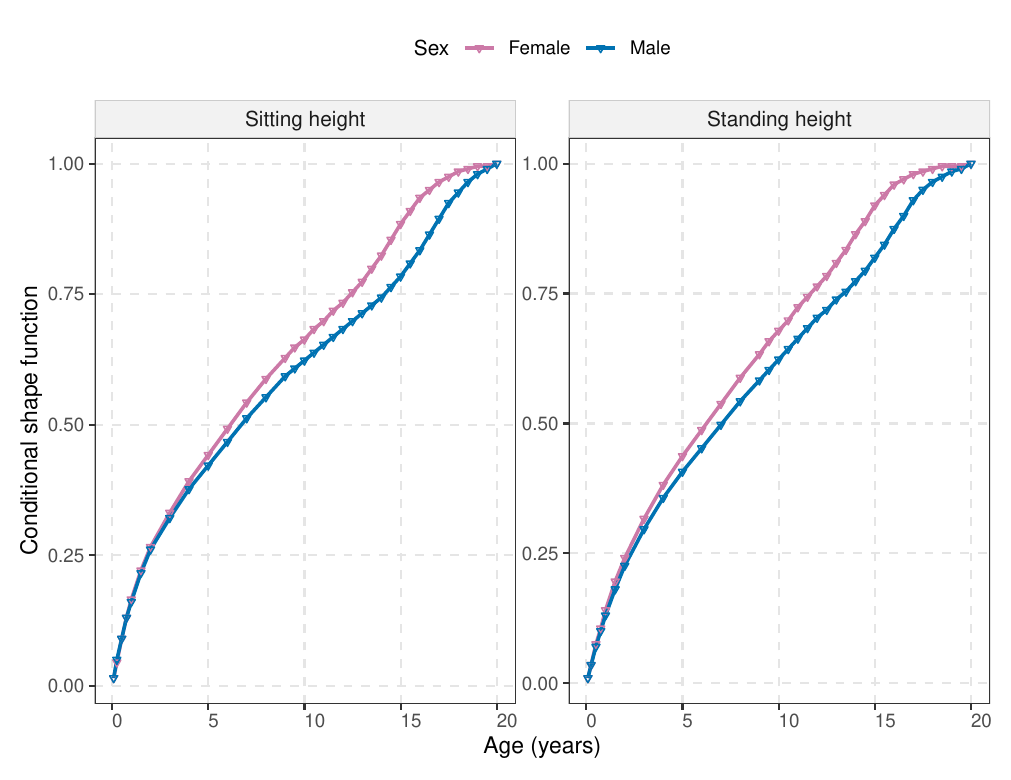}
    \caption{Z\"{u}rich longitudinal growth study with age-varying sitting (left) and standing height trajectories viewed as monotone functional data. Conditional shape trajectories (CDF) obtained for females (pink) and males (blue), using decomposition-based global Fr\'{e}chet regression model.} 
    \label{Zurich:shape:sex}
\end{figure}

\begin{figure}[H]
    \begin{center}
 \includegraphics[height = 8cm, width = 0.8\linewidth]
 {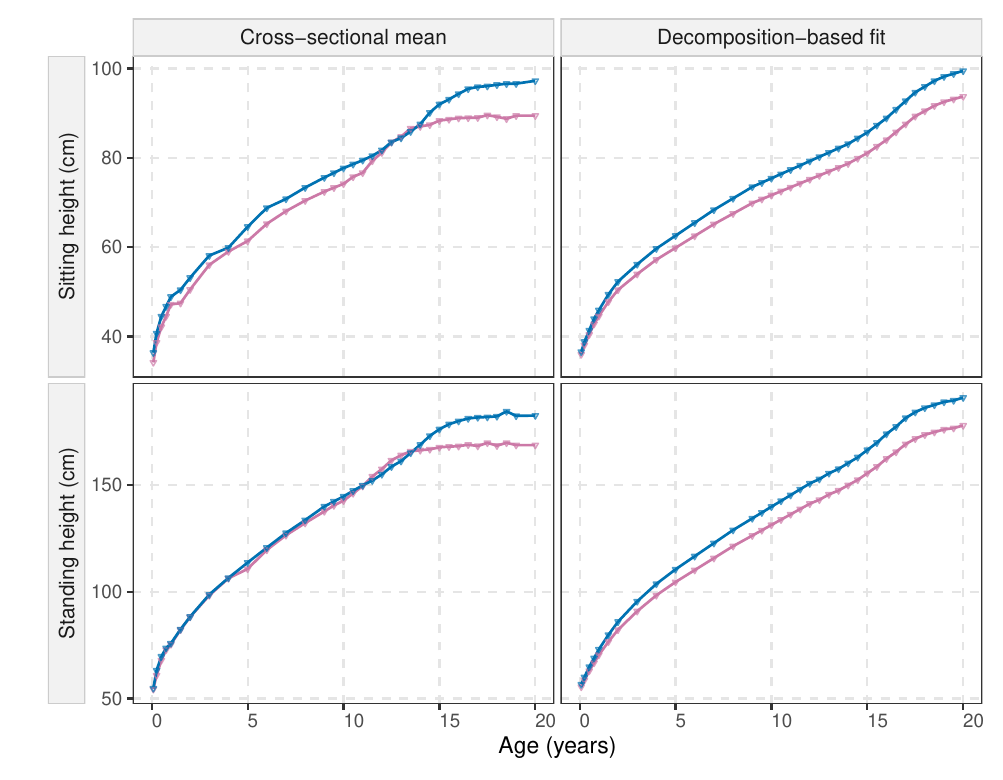}
    \end{center}
    \caption{(Top panel) sitting height and (bottom panel) standing height. Cross-sectional mean height trajectories (left) and conditional height trajectories (right) for females (pink) and males (blue), where the latter are obtained using decomposition-based global Fr\'{e}chet regression model.}
    \label{Zurich:2}
\end{figure}

\begin{figure}[H]
    \begin{center}
 \includegraphics[height = 9cm, width = 0.85\linewidth]
 {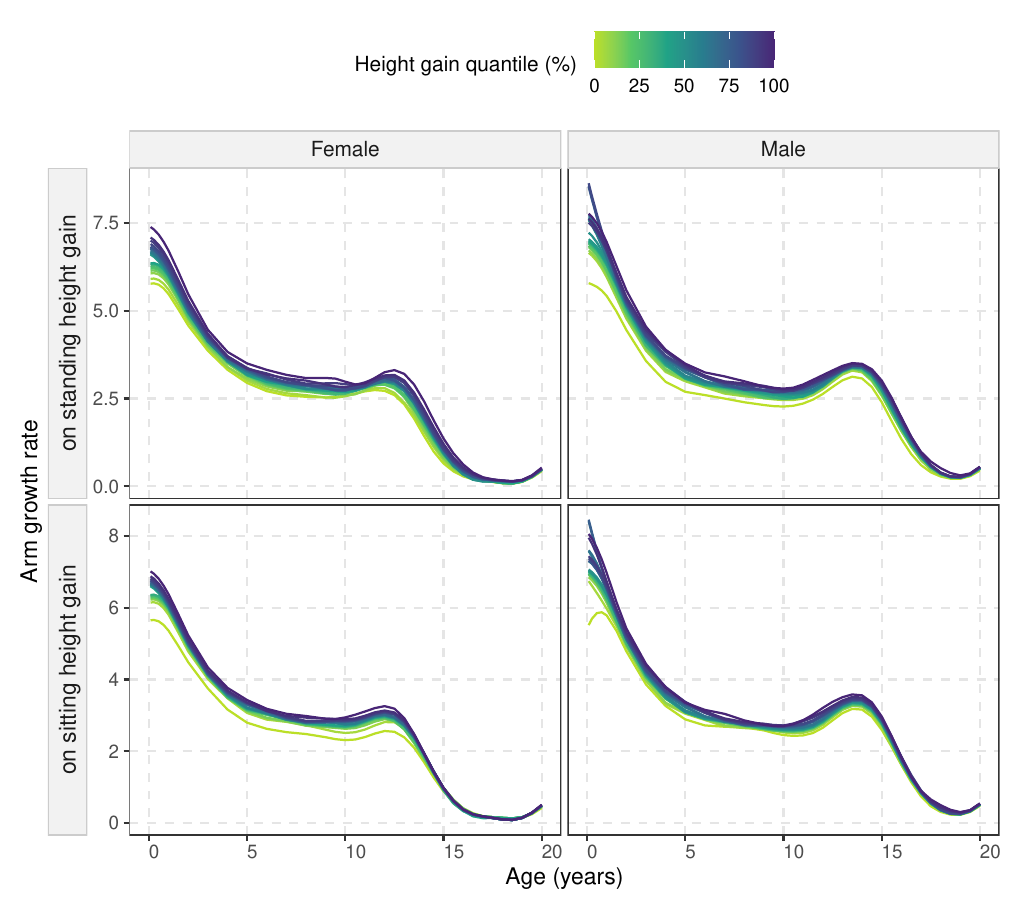}
    \end{center}
     \caption{Z\"{u}rich longitudinal growth study with age-varying arm length trajectories viewed as monotone functional data. Derivatives of fitted arm length trajectories representing corresponding age-varying arm growth rates. Conditional arm growth rate trajectories across varying quantiles of standing (or sitting) height gain for females and males.}
    \label{Zurich:3}
\end{figure}

\end{appendix}

\section*{Funding}
This research was supported in part by NSF grants DMS-2014626 and DMS-2310450.

\bibliographystyle{chicago}
 \bibliography{citation}

\end{document}